\documentclass[preprint,12pt]{elsarticle}

\usepackage{amsmath}
\usepackage{algorithm} 
\usepackage{algpseudocode}
\usepackage{amsmath,amssymb,amsfonts}
\usepackage{algpseudocode}
\usepackage{graphicx}
\usepackage{textcomp}
\usepackage{placeins}
\usepackage{booktabs}
\usepackage{url}
\usepackage{wrapfig,colortbl}
\usepackage{float}

\journal{arXiv.org}

\begin{document}

\begin{frontmatter}



\title{A Roadmap to Holographic Focused Ultrasound
Approaches to Generate Thermal Patterns}


\author[label1]{Ceren Cengiz}
\author[label1]{Zekeriya Ender Eger}
\author[label1]{Pinar Acar}
\author[label2]{Wynn Legon}
\author[label1, corref]{Shima Shahab}
 \ead{sshahab@vt.edu}

\affiliation[label1]{organization={Department of Mechanical Engineering, Virginia Tech},
            city={Blacksburg},
            postcode={24060}, 
            state={VA},
            country={USA}}
            
\affiliation[label2]{organization={Fralin Biomedical Research Institute, Virginia Tech},
            city={Roanoke},
            postcode={24016}, 
            state={VA},
            country={USA}}

\begin{abstract}

In therapeutic focused ultrasound (FUS), such as thermal ablation and hyperthermia, effective acousto-thermal manipulation requires precise targeting of complex geometries, sound wave propagation through irregular structures and selective focusing at specific depths. Acoustic holographic lenses (AHLs) provide a distinctive capability to shape acoustic fields into precise, complex and multifocal FUS-thermal patterns. Acknowledging the under-explored potential of AHLs in shaping ultrasound-induced heating patterns, this study introduces a roadmap for acousto-thermal modeling in the design of AHLs. Three primary modeling approaches are studied and contrasted using four distinct shape groups for the imposed target field. They include pressure-based (BSC-TR and ITER-TR), temperature-based (IHTO-TR), and machine learning (ML)-based (GaN and Feat-GAN) methods. Novel metrics including image quality, thermal efficiency, control, and computational time are introduced, providing each method's strengths and weaknesses. The importance of evaluating target pattern complexity, thermal and pressure requirements, and computational resources is highlighted for selecting the appropriate methods. For lightly heterogeneous media and targets with lower pattern complexity, BSC-TR combined with error diffusion algorithms provides an effective solution. As pattern complexity increases, ITER-TR becomes more suitable, enabling optimization through iterative forward and backward propagations controlled by different error metrics. IHTO-TR is recommended for highly heterogeneous media, particularly in applications requiring thermal control and precise heat deposition. GaN is ideal for rapid solutions that account for acousto-thermal effects, especially when model parameters and boundary conditions remain constant. In contrast, Feat-GaN is effective for moderately complex shape groups and applications where model parameters must be adjusted.

\end{abstract}


\end{frontmatter}



\section{Introduction}
\label{sec1}

The intersection of acoustic and thermal phenomena within the domain of ultrasound technology has revealed the heating potential of sound waves when strategically harnessed. This acousto-thermal energy transformation has found applications in various fields, including industrial operations, material characterization, smart-structure actuation, and therapeutic ultrasound \cite{ku2001applications, zohm2000thermal, ibarra2009comparative, bhargava2019coupling, baghban2017estimation, escoffre2015therapeutic}. Notably, advanced control of ultrasound-induced thermal energy is critical for the successful implementation of these applications. This control encompasses but is not limited to, triggering shape-memory polymers of diverse sizes and geometries \cite{zhu2024shape}, ensuring accurate targeting with precise thermal control for the safe treatment of complex and irregular structures \cite{xu2024safety}, and generating spatially confined damage to prevent surrounding cellular damage for selective destruction  \cite{ter2007high}. 

Integrating an acoustic holographic lens (AHL) with a flat transducer is one method for obtaining advanced control over ultrasound-induced thermal energy. Compared to the widely employed curved or phased-array transducers, AHLs have proven their potential for their cost-effectiveness, robustness, and ability to offer higher resolutions \cite{melde2016holograms, brown2020stackable, zhang2020acoustic, jimenez2019holograms,sallam2021holographic, bakhtiari2018acoustic}. While the majority of applications utilizing AHLs for sound manipulation have been focused solely on precise control of pressure fields, recent years have seen a growing recognition of their potential to induce heating effects \cite{andres2022thermal, andres2023holographic, li2022generating} and the necessity to produce AHLs specifically for thermal applications \cite{kim2023nanoparticle}.

The underlying principle behind AHLs lies in their ability to function as phase plates, modulating the phase of the wavefront by employing a unique thickness map. By adjusting the thickness and speed of sound in the design, the uniform pressure field distribution of the flat transducer can be transformed into arbitrary pressure patterns at the desired location. AHLs are often manufactured using high-resolution 3D printing methods, such as stereolithography (SLA). This 3D production process simplifies the conversion of the generated thickness map into a physical holographic lens design.
 
Obtaining the necessary thickness map for 3D production is the primary challenge in utilizing AHLs, which largely depends on numerical analysis. Various algorithms have emerged for modeling wave propagation across diverse domains \cite{gu2015modeling} and have been utilized in generating thickness maps for AHLs \cite{fushimi2021acoustic, sallam2023nonlinear, ferri2019enhanced, jimenez2021acoustic, sallam2024gradient}. Additionally, discussions concerning homogeneous linear propagation and the appropriateness of modeling approaches for heterogeneous media have garnered attention from the community and are ongoing, with room for further improvement. However, the exploration of incorporating acousto-thermal conversion into AHL design and the expansion of the application scope for creating ultrasound-induced heat patterns remains relatively under-explored. 

This paper introduces a comprehensive road map for acousto-thermal modeling in the design of thickness maps for AHLs. By incorporating both established and novel approaches, we specifically created and analyzed thickness maps to effectively impose thermal patterns. Although our primary focus is on time-reversal based full wave acoustical modeling, the methodologies described are adaptable to other modeling techniques. The paper begins by reviewing two acoustical-based techniques that involve forward and backward propagation of the pressure fields for acoustic pattern generation. Following that, a strategy for addressing the inverse heat transfer problem (IHTP) to directly induce desired thermal patterns to the design procedure of the AHL is explored. Additionally, two novel machine learning-based approaches are introduced in the methodology section, aiming to streamline the multi-physics aspects into a single inverse problem framework while minimizing computational effort. One of the ML algorithms is tailored to provide a case-specific solution for the design of AHL, while the other is a generic model augmented with additional parametric features. In the results and discussion sections, each method is compared, highlighting their respective advantages and disadvantages under various considerations. A new quality metric is introduced to enable a fair comparison of findings, accompanied by a detailed analysis of imposed target pattern complexity to distinguish geometric features. Potential clinical applications of AHL assisted focused ultrasound (FUS) are also discussed in the results and discussion section to offer application directions to the readers. Ultimately, this paper provides a comprehensive analysis of the strengths and weaknesses of various computational methods for AHL design, guiding the selection of the most suitable technique for different FUS applications.

\section{Methodology}
\label{Method}

Time reversal (TR) techniques in acoustics offer an effective means of focusing the sound waves emitted from a source \cite{fink1992}. Fundamentally, TR techniques leverage the special temporal characteristics to solve the second-order governing equation for sound propagation \cite{kinsler2000fundamental}:

\begin{equation}
\nabla^2p-\frac{1}{c^2}\frac{\vartheta^2p}{\vartheta t^2}
\label{eq:wave}
\end{equation}

Where \(p\) represents the acoustic pressure and \(c\) is the speed of sound. The inclusion of the second-order time derivative in Eq. \ref{eq:wave} enables both \(p(t)\) and \(p(-t)\) to function as solutions to the pressure field in the wave equation \cite{fink1999time}. Consequently, this allows a forward-propagated acoustic wave to be traced back to its source by simply reversing the recorded time-varying pressure. While acoustic attenuation in water and similar mediums remains relatively low enough for Eq.\ref{eq:wave} to hold true, the introduction of an attenuating secondary medium, such as biological tissue, necessitates additional terms in the governing equations. The generalized Westervelt equation for acoustic propagation, as provided in Eq.\ref{eq:westervelt}, encompasses these acoustic energy losses and nonlinearities.

\begin{equation}
    \nabla^2p -\frac{1}{c^2}\frac{\partial^2p}{\partial t^2}+\frac{\delta}{c^4}\frac{\partial^3p}{\partial t^3}+\frac{\beta}{\rho c^4}\frac{\partial^2p^2}{\partial t^2} = 0
    \label{eq:westervelt}
\end{equation}

Acoustic energy losses are included in the third term of the equation where \(\delta\) represent the sound diffusivity. The same term also includes the third-order time derivative, which can challenge the assumption of time invariance. If the attenuation coefficient remains low enough for the studied frequency range, time invariance can be assumed to be valid for biological mediums \cite{fink1992}, and TR can still be successfully employed. However, in other cases, attenuation compensation should be considered for utilization of TR techniques \cite{treeby2013acoustic}.

In low-pressure regimes, the nonlinearity term (\(\beta\)) in Eq.\ref{eq:westervelt}   can be disregarded. However, when high intensities are necessary for inducing ablation, nonlinear distortions should be accounted for generation of AHLs \cite{sallam2022nonlinear} and for enhancing the efficiency of the acoustic heating process \cite{filonenko2001effect}. The presented study will primarily exclude nonlinearity terms from computations, especially during general method comparisons where temperature rises are relatively low compared to highly nonlinear pressure regimes. However, necessary discussions will still be provided for the incorporation of these effects.

After solving the Westervelt equation to obtain acoustic pressure within the target domain, heat generation per unit volume (\(Q\)) is calculated using the Eq.\ref{eq:heat deposition} to study the acousto-thermal effects.

\begin{equation}
Q= \frac{\alpha\\p^2}{\rho c }
\label{eq:heat deposition}
\end{equation}

Where \(\rho\) stands for the density and the parameter \(\alpha\) represents the absorption coefficient, indicating the loss of wave energy during propagation. Its calculation follows the following frequency power law: 

\begin{equation}
\alpha=\alpha_0\\f^\gamma\  
\label{eq:absorption}
\end{equation}

A power law exponent value of \(\gamma \cong 1\)  has been demonstrated to be a suitable approximation for the frequency-dependent (\(f\)) attenuation model of soft biological tissues  \cite{parker2022power}. As the absorbed acoustic field is converted into a heat deposition, the resulting thermal variation needs to be evaluated as a final step. 

Due to simplicity and straightforwardness, Pennes' bioheat approximation \cite{pennes1948analysis} is widely employed for heat transfer modeling of biological tissues.

\begin{figure*}
    \centering
    \includegraphics[width=1\linewidth]{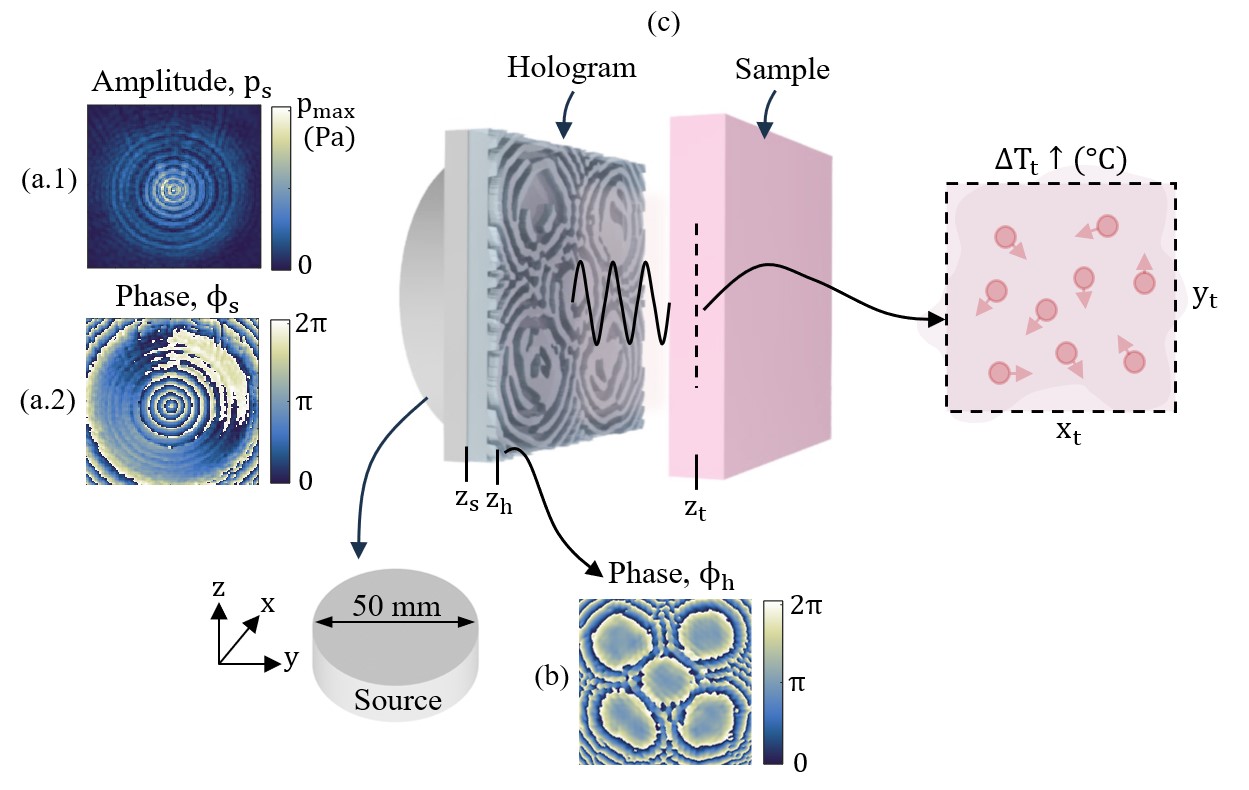}
    \caption{Overview of the methodology for generating AHL-induced heating effects. (a.1) Amplitude map (\(p_{s}\)) at the source plane (\(z_{s}\)), (a.2) Phase map (\(\phi_{s}\)) at the source plane, (b) Hologram plane (\(z_{h}\)) phase map illustrating the phase modulation of the acoustic wavefront, (c) Schematic of the model domain and acousto-thermal conversion.}
    \label{fig:enter-label}
\end{figure*}

\begin{equation}
\rho C \frac{\partial T}{\partial t} = \nabla \cdot (\kappa \nabla T) + Q_m + Q - Q_b   \label{eq:bioheat}
\end{equation}

Where \(T\) is the temperature, \(\kappa\) is thermal conductivity and \(C\) is the specific heat. In addition, heat deposition by the acoustic absorption, heat generated by the metabolism (\(Q_m\)) and blood perfusion effect (\(Q_b\)) can also be included in the general form of the Eq.\ref{eq:heat deposition}. Notably, efforts have been made to enhance Pennes' Bioheat equation \cite{chen1980microvascular,yang2007expanding, yuan2008numerical,LIU20101138, gupta2019non} by addressing accuracy issues stemming from assumptions such as uniform perfusion rate, omission of intricate anatomical characteristics, and non-Fourier conduction behavior, among others. While these efforts have shown promising improvements in the accuracy of heat transfer models for biological media, their complexity makes it challenging to implement the anatomical features and determine the necessary parameters \cite{andreozzi2019modeling, singh2020thermal}. Therefore, as of now, Eq. \ref{eq:bioheat} continues to be widely utilized as a prevalent modeling approach for general applications of heat transfer analysis in biological systems.

To provide a brief summary of acoustic holographic lens generation, TR simulations in the full-wave time domain are conducted by first placing virtual sources to the target regions in the desired shape within the target medium (\(z_t\) in Fig.\ref{fig:enter-label}). The ultrasound field is then backpropagated to the hologram plane (\(z_h\) in Fig.\ref{fig:enter-label}) transmitted through the target and propagation mediums. The acoustic wave signal is recorded by the sensors distributed at the \(z_h\) and the Fourier transform is performed to obtain the necessary phase \(\phi_h\) and/or amplitude \(p_h\) information at the \(z_h\). This information is then used to compute the thickness map matrix (refer to Eq.\ref{thicknessmap}) and is converted to a Standard Tessellation Language (STL) format for 3D printing the AHL \cite{melde2016holograms}.

\begin{equation}
\label{thicknessmap}
s_{(x,y)}=s_0-\frac{\Delta\phi_{(x,y)}}{k_w-k_h}
\end{equation}

Here, each pixel's thickness at the \(z_h\) plane denoted as \(s_{(x,y)}\), while the phase difference is expressed by \(\Delta\phi_{(x,y)}\). The initial thickness of the hologram plate is represented by \(s_0\) and the wave numbers for the hologram and surrounding medium is given by \(k_h\) and \(k_w\), respectively.

In this paper, k-space pseuodospectral method is used for computations to obtain a time domain solution of the wave equation in heterogeneous media \cite{kwave}. In all simulations, the computational domain is formed with 300x300x130 grid points in X, Y, and Z directions respectively. The spatial resolution is set to 0.2 mm size which corresponds to \(\lambda/6\) in water at the selected driving frequency of 1 MHz.  This excitation frequency aligns with the common range used in therapeutic ultrasound applications \cite{speed2001therapeutic}. Perfectly matched layers (PMLs) are also incorporated into the computational domain to eliminate reflections at the boundaries. 
For all cases, water is chosen as the primary propagating medium, while the properties of the secondary medium are adjusted to mimic those of soft brain tissue. As reported in \cite{softtissue}, acoustical properties of the water and soft tissue are defined as: \(c_w = 1480\)  m/s, \(\rho_w = 1000\)  kg/m$^3$, \(\alpha_w = 0.0022 \) dB/cm.MHz, \(c_t = 1560 \) m/s , \(\rho_t = 1040\) kg/m$^3$,  \(\alpha_t = 0.6 \) dB/cm.MHz, respectively. For heat transfer calculations, specific heat and thermal conductivity of the brain tissue is defined as \(C=3630\) J/kg.K and \(\kappa=0.51\) W/m.K \cite{thermalprop}. The initial temperature of the tissue and the temperature at the model boundaries are set to 37°C. Experimentally measured hologram plane pressure field is applied as the boundary condition (see \(p_s\) and \(\phi_s\) in Fig. \ref{fig:enter-label}) incorporating actual acoustic source data into the computations. To achieve this, pressure mapping experiments are conducted using a needle hydrophone (ONDA HNR-0500) controlled by a 3D positioning system. For more details on experimental measurement for source characterization, readers can refer to the previous study \cite{10.1088/1361-6463/ad5452}. Once the acoustic source is experimentally characterized, pressure field data for both forward and backward propagation are enforced as Dirichlet boundary conditions in the simulations. Open source K-wave MATLAB toolbox is used for the numerical analysis \cite{treeby2010k}. The simulations utilized a high-performance computing node on Virginia Tech's Infer cluster, with 28 cores (Intel Xeon E5-2680v4 2.4GHz CPUs), 512GB of RAM, and 187GB SSD.

\subsection{Pressure based approaches}
\subsubsection{Basic Time Reversal (BSC-TR)}
\label{sec:BSC-TR}

The basic time reversal (BSC-TR) approach, as the name suggests, serves as the baseline for the time reversal technique. In this method, the desired thermal field is assumed to match the distribution of the resulting pressure amplitude (\(p_t\)) and imposed by placing monopole virtual sources at the \(z_t\). Sinusoidal signals emitted from these source points are then backpropagated to the \(z_h\). The recorded time-varying source field is subsequently processed to extract the \(\phi_h\) and \(p_h\) information required for generating the thickness map. If no post-processing is done on the obtained \(\phi_h\) field, the next step is to directly forward propagate the \(\phi_h\)  with experimental boundary conditions to obtain the final \(p_t\). An important shortcoming of BSC-TR is that the \(p_h\) is time-reversed in an uncontrolled manner and extracted data does not necessarily match the amplitude field of the actual source. Since only the phase field can be modulated with AHL, discrepancies between the forward-propagated and initially imposed pressure fields are expected. One way to overcome this challenge is by applying error diffusion schemes to convert the complex acoustic field into a phase-only distribution. The Bidirectional Error Diffusion (BERD) method, originally introduced for Fresnel holograms \cite{tsang2013novel}, has been successfully utilized for generating phase-only acoustic holograms \cite{jimenez2019holograms} using time reversal methods. Recently, an improved version of the BERD algorithm has been suggested by incorporating a pattern-adaptive amplitude coefficient to replace the constant value of 1 \cite{liu2021pattern}. A comparable methodology is also employed in this study where a phase-only hologram is generated by adjusting the amplitude coefficient based on the target shape. Equations from \cite{tsang2013novel} are adapted for unidirectional and bidirectional error diffusion in odd and even row pixels, respectively. The main distinction of pattern-adaptive BERD algorithm lies in the error equation (Eq. \ref{eq:H_POH}) applied to determine the corresponding error value. Where the value \(m\) is used to regulate the extent to which the amplitude field is integrated into the phase field. \(H\) is the complex hologram field formed by the \(\phi_h\) and \(p_h\) while the \(H_{POH}\) is the phase-only hologram. Details on the variation of the \(H_{POH}\) with respect to amplitude parameter \(m\) is given in \ref{berd} as a reference.

\begin{equation}
E(i,j)= H(i,j) - m\times H_{POH}(i,j) 
\label{eq:H_POH}
\end{equation}

Up to this step of BSC-TR, thermal effects had no influence over the generated \(H_{POH}\), and the \(\phi_h\) obtained via BSC-TR does not account for the variations that can be introduced during the heat transfer process. To calculate the resulting temperature rise, first, the acousto-thermal conversion is done using the relation given in Eq.\ref{eq:heat deposition}, followed by the calculation of the resulting temperature increase via Eq.\ref{eq:bioheat}. The magnitude of the pressure amplitude or the duration of heating can be adjusted to achieve the desired temperature rise. However, this process requires a time-consuming series of trial-and-error iterations until the exact thermal conditions are reached.

\subsubsection{Iterative Time Reversal (ITER-TR)}

Iterative approaches have been recognized for their success in generating AHLs \cite{melde2016holograms, bakhtiari2018acoustic, sallam2021holographic}, particularly in cases where rapid simulation methods such as the angular spectrum approach (ASA) are viable for acoustic wave propagation in a homogeneous medium. Similarly, iterative time reversal (ITER-TR) aims to enhance the reconstruction of the pressure field obtained with BSC-TR by imposing experimentally obtained values of \(\phi_t\) and \(p_h\) from the acoustic source. Although the number of iterations will be limited due to computational time constraints of full-wave simulations in heterogeneous mediums, ITER-TR is expected to enhance the reconstructed \(p_t\) and, consequently, the corresponding temperature map.

The ITER-TR method begins by forward propagating the source field from \(z_s\) plane to both \(z_h\) and \(z_t\) planes to extract the initial phase and amplitude distributions at the two primary planes of interest. Next, the complex acoustic field formed by the desired \(p_t\) and actual \(\phi_t\) is back propagated to the \(z_h\) plane. While the obtained \(\phi_h\) remains unchanged, experimentally measured \(p_h\) is imposed with the \(\phi_h\) to form the new force field and is forward propagated to the \(z_t\) plane. Utilizing Eq. \ref{eq:heat deposition} and Eq.\ref{eq:bioheat}, the temperature distribution at \(z_t\) is assessed. If the thermal pattern at the target plane is deemed unsatisfactory, the desired \(p_t\) is reapplied, while the forward propagated \(\phi_t\) field from the previous step is kept the same. This process of back-and-forth propagation of the acoustic field continues until satisfactory results are achieved. The primary objective is to maintain constant values for the source field amplitude and target plane pressure field while allowing the phase distribution to vary in each iteration. Similar to BSC-TR, heat transfer effects do not impact the fluctuating phase field in each iteration.
One critical aspect of ITER-TR is determining the termination criteria for the iterations. Given the time-consuming nature of full-wave simulations, a performance criterion is necessary to determine the optimal number of iterations that yield satisfactory results within a reasonable computational time frame. 

Peak Signal-to-Noise Ratio (PSNR) and Structural Similarity Index (SSIM) are well-known image quality metrics used to assess the performance of image reconstruction relative to the ground truth image. Additionally, \(T_{\mathrm{eff}}\) is defined to evaluate thermal reconstruction efficiency.

\begin{equation}
\label{eq:PSNR}
PSNR= 10log_{10}\frac{(MAX)^2}{MSE}  
\end{equation}
\begin{equation}
\label{eq:MSE}
MSE = \frac{1}{ab} \sum_{i=1}^{a} \sum_{j=1}^{b} [\hat{\Delta T}(i,j) - \Delta T(i,j)]^2
\end{equation}
\begin{equation}
\label{eq:SSIM}
SSIM =  \frac{{(2\mu_x\mu_y + L_1)(2\sigma_{xy} + L_2)}}{{(\mu_x^2 + \mu_y^2 + L_1)(\sigma_x^2 + \sigma_y^2 + L_2)}}
\end{equation}
\begin{equation}
\label{eq:Teff}
 T_{eff}=  \frac{\sum (\Delta T(i_{in},j_{in})^2}{\sum (\Delta T(i_{in},j_{in})^2 +\Delta T(i_{out},j_{out})^2)}
\end{equation}

The term \textit{MAX} denotes the maximum pixel value, which can be determined using the expression \(2^D-1\), where \textit{D} represents the dynamic range of the ground truth thermal image. The \textit{MSE} calculations, as given in Eq.\ref{eq:MSE}, involve \(\hat{\Delta T} \) and \(\Delta T\) representing the ground truth thermal pattern and simulated thermal pattern, respectively. The '\textit{i}' and '\textit{a}' notations denote the rows, while '\textit{j}' and '\textit{b}' denote the columns of the data matrix. In Eq. \ref{eq:SSIM}, local mean values (\(\mu_{x}\) and \(\mu_{y}\)), standard deviations (\(\sigma_x^2\) and \(\sigma_y^2\))  and stability constants (\textit{L1} and \textit{L2}) are utilized to calculate the SSIM. Lastly, for the \(T_{\mathrm{eff}}\) in Eq. \ref{eq:Teff}, subscripts '\textit{in}' and '\textit{out}' are defined to represent rows and columns with \(\Delta T > 0\) and  \(\Delta T = 0\), respectively. Although these metrics offer guidance for terminating simulations once satisfactory results are achieved, the application of such criteria varies. For example, there isn't a universal threshold value for PSNR to guarantee optimal high-quality reconstruction. Similarly, for SSIM and \(T_{\mathrm{eff}}\), values closer to one signify higher quality, however, termination criteria should be set below this to avoid excessive iterations or unattainable targets. Each metric offers distinct insights, and a singular endpoint criterion may not adequately reflect all quality aspects. Calculating a performance metric that considers various quality metrics simultaneously could be a way to address the issue, as discussed in detail in Section \ref{ResultsandDisc}.

\label{sec:ITER-TR}
\FloatBarrier

\subsection{Temperature based approach}
\subsubsection{Inverse Heat Transfer Optimization (IHTO-TR)}
\label{sec:IHTO-TR}

Inverse heat transfer problems (IHTP) are known for their ill-posed nature \cite{ozisik2018inverse}. In most cases, IHTPs are inherently unstable and highly sensitive to input data, presenting significant challenges in practical applications. Despite these difficulties, considerable efforts have been devoted to developing solutions for IHTP \cite{taler1999solution, weber1981analysis, alifanov2012inverse, su2004inverse}, due to the critical need for accurate and reliable thermal solution methods in various engineering applications.

In both the BSC-TR and ITER-TR methods mentioned above, determining the acoustic field at the hologram plane involves an inversion process that primarily focuses on the pressure amplitude distribution. After solving the inverse problem, the analysis shifts to heat transfer, where Eq.\ref{eq:heat deposition}-Eq.\ref{eq:bioheat} are utilized in a forward manner to calculate the final \(\Delta T\) within the target medium. This approach overlooks the importance of the thermal field, which is actually left uncontrolled. Without incorporating heat transfer dynamics into the problem, discrepancies arise between the desired and obtained thermal fields. To convert the problem into an inverse one that considers relevant heat transfer effects, the solution of the IHTP should be integrated into the numerical model. 

In this section, an Inverse Heat Transfer Optimization (IHTO) is combined with the TR technique to account for the acousto-thermal effects when generating the AHLs. For this purpose, we leverage the conjugate gradient method (CGM), a successful iterative regularization technique widely utilized in various IHTPs \cite{huang1999three, perakis2019inverse, haghighi2008two} and in recent years, it has been employed for solving inverse BioHeat relations for thermal therapy studies \cite{baghban2017estimation, lee2013inverse, mehrabanian2023new}. An overview of the CGM method and its equations are provided in Algorithm 1, where the initialization parameters include $\eta$ as the learning rate, $\epsilon$ as the convergence criteria, and $\beta$ as the search step size. Moreover, the descent direction in the CGM is denoted by $d$, and $r$ represents the gradient of the objective function. As demonstrated in Algorithm 1, an adaptive learning rate is utilized to dynamically adjust the learning rate during the optimization process, ensuring smoother convergence and preventing overshooting. To ensure that predicted values of heat deposition remain positive, the algorithm enforces zero for any negative values. In summary, the algorithm steps are as follows (1) Evaluate  \(Q\) by solving IHTP with CGM. (2) Solve the direct problem for the heat transfer analysis and determine the resulting temperature field (3) Calculate the error between \(T_{\text{predicted}}\) and \(\hat{T}\) (4) Repeat steps 1-3 until the desired convergence criterion is met.

\begin{algorithm}
\caption{Conjugate gradient method for IHTO}\label{algorithm}
\begin{algorithmic}[1] 
    \State \textbf{Result}: Heat deposition $Q$
    \State \textbf{Initialization:} Initial field configuration $Q_0 = |\hat{T}|$
    \State \textbf{Initialization:} $\eta_{1}=0.5$, $\eta =\frac{\eta_{n}}{5}$, $n=1,2,3$
    \State \textbf{Initialization:} $\epsilon =0.1$, $\beta = 0.6$
    \While{not converged}
        \State $r = 2(T_{\text{predicted}}-\hat{T})$
        \State $d^{k} = r^{k}+ \beta d^{k-1}$
        \State Update heat source: $Q_{\text{predicted}} = Q_{\text{predicted}} -\eta d^{k}$
        \State $Q_{\text{predicted}}(Q_{\text{predicted}}<0)=0$ \Comment{Ensure non-negativity}
        \State Calculate current cost: $\epsilon_{\text{calculated}} = \frac{1}{ab} \sum_{i=1}^{a}\sum_{j=1}^{b} |r_{i,j}|$
        \If{$\epsilon_{\text{calculated}}<2\epsilon$}
            \If{$\epsilon_{\text{calculated}}<1.5\epsilon$}
                \State $\eta = \eta_{3}$
            \Else
                \State $\eta = \eta_{2}$
            \EndIf
        \EndIf
        \If{$\epsilon_{\text{calculated}} \leq \epsilon$}
            \State \textbf{break}
        \EndIf
    \EndWhile
\end{algorithmic}
\end{algorithm}

After computing \(Q\) using CGM, Eq. \ref{eq:heat deposition} is employed to determine \(p_t\) within the target medium. Subsequently, TR techniques are applied to ascertain the acoustic field at \(z_h\). Depending on preference, CGM can be integrated with either BSC-TR or ITER-TR, as TR assessments and inverse heat transfer calculations occur separately. Additionally, it is possible to amalgamate IHTO with forward and backward propagations of the TR simulations. The descent direction and gradient parameters derived from the merged IHTO and TR methods may be more relevant for the actual forward-propagated temperature field. This approach could potentially lead to more suitable parameters, thereby facilitating faster convergence. Yet, the computational expenses of TR, necessitate adjustment with adopted optimization technique to expedite convergence with fewer iterations. Otherwise, merged TR and IHTO should be considered for scenarios incorporating other sound propagation techniques or heat transfer inversion methods, as full wave simulations will be computationally demanding to be combined with iterative regularization techniques to solve the inverse problem. One possible suggestion could be to utilize non-gradient-based optimization methods, which may offer computational advantages as they do not rely on derivatives. These methods have shown success in multiple studies for inversion of the BioHeat equation \cite{paruch2007identification, majchrzak2011identification, hossain2016thermogram}. It is important to consider that solution accuracy and computational efficiency are not always guaranteed, especially given the potential complexity of the direct problem \cite{scott2018inverse}. Overall, the advantage of IHTO-TR over BSC-TR and ITER-TR lies in its ability to incorporate heat transfer dynamics centrally within the computations. Considering the success of IHTO solutions in determining the source term in a number of therapeutic applications for thermal dose optimization \cite{loulou2002thermal, kuznetsov2006optimization, baghban2015source, liu2020optimized, mehrabanian2023new}, IHTO can offer better applicability over BSC-TR or ITER-TR for future real-life applications of AHL assisted generation of FUS. 

\subsection{Machine-learning based approaches}
\subsubsection{Generative adversarial network (GaN)}
\label{sec:GaN}

In addressing the challenges associated with incorporating heat transfer-related effects into AHL design, machine learning-based (ML) approaches emerge as promising solutions. ML advancements can facilitate AHL design optimization by enabling precise thermal control to achieve the desired target thermal pattern with enhanced computational efficiency. ML-assisted AHL generation has been explored in a few studies \cite{lin2021deep, lee2022deep, li2022acoustic}. Yet its application in heterogeneous domains to solve the IHTP and TR techniques combined has not yet been thoroughly investigated. The aim of incorporating the ML approaches introduced here is to address the entire problem by consolidating it into a single inverse problem, where acousto-thermal conversions are resolved collectively within the proposed framework. 

Generative Adversarial Networks (GANs) is one of the classes in deep learning that consist of two neural networks, the generator and the discriminator \cite{goodfellow2014generative}. The generator aims to create realistic data, such as images, audio, or text, while the discriminator's role is to distinguish between real and generated data. The success of the generator is measured in its ability to deceive the discriminator, such that the discriminator will not be able to identify the difference between real and generated data. Through adversarial training, GANs learn to generate increasingly convincing outputs, often achieving results in image generation, style transfer, and data augmentation. Throughout the years, many developments have been suggested for the GAN algorithm \cite{creswell2018generative,durgadevi2021generative}. Examples include the introduction of conditional GANs, which allow for the generation of data conditioned on specific input information, leading to applications like image-to-image translation \cite{mirza2014conditional}. The nature of the problem poses similarity to the ones that the pix2pix algorithm worked well on, mainly tasks such as image-to-image translation, where a model learns to map input images from one domain to output images in another \cite{isola2017image}. The pix2pix algorithm utilizes a conditional GAN framework, combining it with a paired image dataset for supervised learning, enabling tasks like converting satellite images to maps, turning sketches into realistic images, or transforming day-time scenes into night-time scenes, among others. The generator uses U-net in the network architecture and utilizes a composite loss function, unlike the discriminator. In addition to the idea of rewarding the generator when it is able to generate realistic images, since the target image is present a Mean Squared Error (MSE) loss is added where the difference between target and generated images is minimized.

\begin{figure*}
    \centering
    \includegraphics[width=1\linewidth]{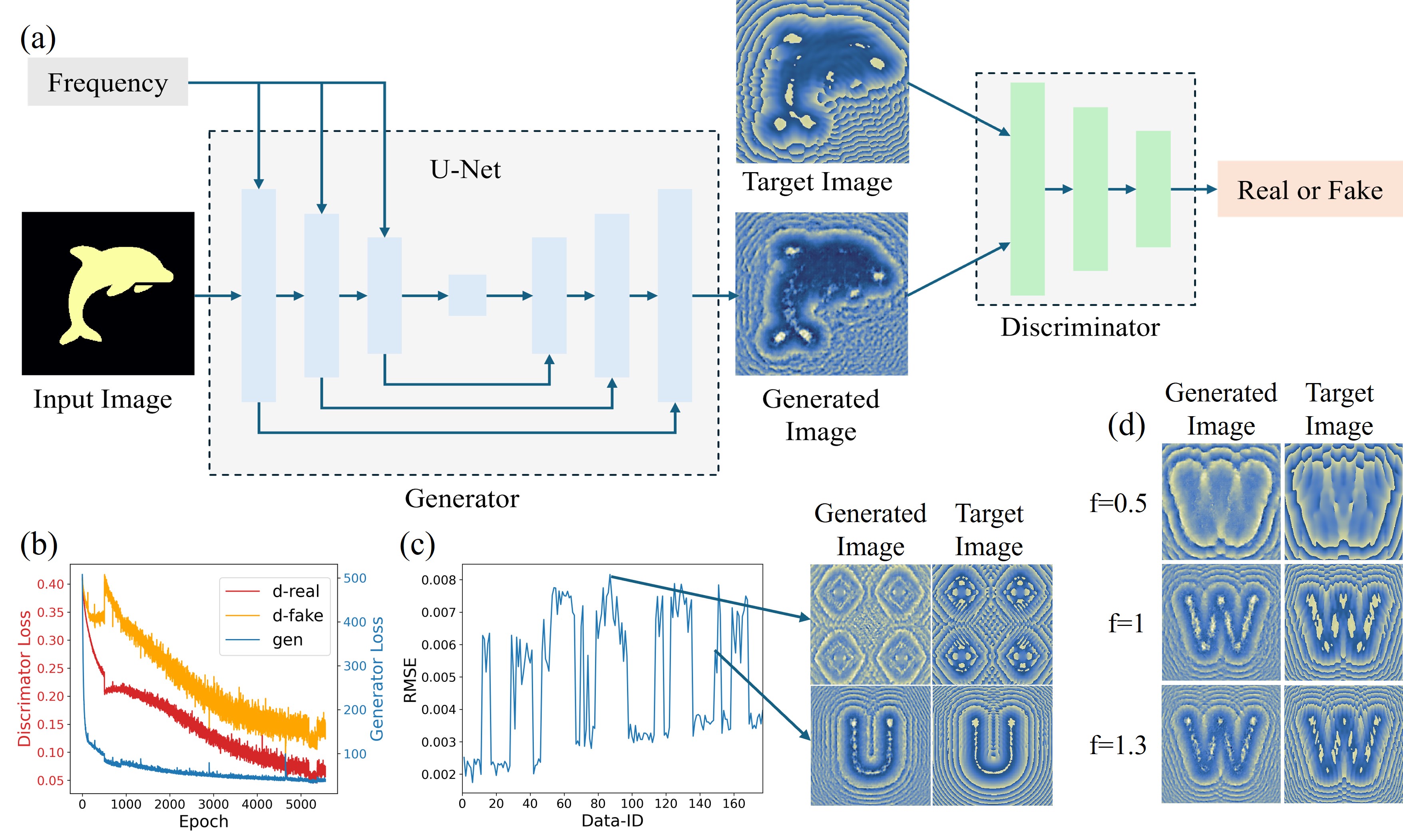}
    \caption{(a) Proposed framework of the pix2pix architecture for frequency-dependent phase and amplitude generation. (b) Loss values for discriminator and generator during training. Discriminator has 2 loss values, loss of predicting real shown in red and loss of predicting fake shown in orange. (c) Evaluation of the trained model across the entire dataset and predictions for testing datasets with highest and lowest RMSE values. (d) Prediction for varying frequencies for an image in the testing dataset.}
    \label{fig:overviewFTGAN}
\end{figure*}
In the GaN method presented in this study, pix2pix algorithm is applied to obtain the \(\phi_h\) and \(p_h\) information required to construct the desired \(z_{t}\) layer temperature distribution. To achieve that temperature map is provided as input for both of the \(\phi_h\) and \(p_h\) field information. Training is performed separately with 78 pairs for each class and the hyperparameters as well as the coefficients for composite loss function comprising both MSE and binary cross entropy loss are set based on the original paper \cite{isola2017image}. The training is completed in 76,800 s on the cluster with an AMD
EPYC 7702 CPU clocked at 3.35 GHz. The method is trained for a case-specific situation where the number of datasets for training is kept low. The aim here is to develop a working ML system with a low number of training data, tailored for a specific scenario where variations in distance, frequency, or acousto-thermal properties are minimal, allowing for full control over temperature and target shape variations within that scenario. Correspondingly, the system fully considers the effect of acousto-thermal gradients within the formation of the target temperature field and is capable of generating hologram plane data suitable for various target shape groups.

\subsubsection{Generative Adversarial Network with Feature (Feat-GAN)}
\label{sec:Feat-GaN}
The pix2pix architecture of the GaN method is extended by integrating additional features, drawing inspiration from the innovative concepts introduced in InfoGAN \cite{chen2016infogan} and StyleGAN \cite{karras2019style}. InfoGAN adds a new layer of complexity to GANs by introducing a latent code vector alongside the noise vector. This latent code, learned in an unsupervised manner, governs specific features of the generated output, enabling the network to learn representations that are hard to capture. Similarly, StyleGAN improves the GAN landscape by introducing a novel style-based synthesis technique. Departing from the conventional latent space, StyleGAN divides the latent space into two components: the traditional latent space and a separate style space. By providing the generator not only with a random noise vector but also with a latent code controlling the style of the generated image, StyleGAN offers control and diversity in data generation.

In this study, these advanced concepts are adapted and simplified for our specific task with the concept of BicycleGAN \cite{zhu2017toward}. In the framework, the additional input, that
is frequency, is integrated directly into the generator network, without any intermediary networks. By incorporating frequency as an additional input feature, we aim to equip the generator with the ability to produce \(z_{h}\) data with varying characteristics. To streamline the training process for the development of a more versatile ML system, temperature maps are converted into binary maps to represent different shapes, with less emphasis on the impact of temperature gradients on the formation of the \(z_{h}\)
layer pressure field data. Nevertheless, the algorithm remains capable of capturing the internal relationships between acousto-thermal conversions as it still addresses the inverse problem with heat transfer phenomena included. However, greater attention is devoted to the effects of various parameters rather than the precise values of temperature elevations. In order to ensure fairness and prevent biases in our training process, the strategy of using full batches and balancing the dataset by providing an equal representation of data for each frequency is adopted. In the composite loss function, the ratio of coefficients of MSE and cross entropy is increased and training is carried out with a much lower learning rate. The data set consists of 176 pairs of \(\phi_{h}\)-\(p_{h}\) and 84 of them are used for training whereas the rest 92 of them are used for testing. The training is performed on the same processor and lasted for 172,860 seconds.
For the training of phase map, which is similar to amplitude training, the loss values of both the discriminator and generator networks, are illustrated in the convergence plots depicted in Fig. \ref{fig:overviewFTGAN}(b). Following training, the trained model's performance is assessed across the entire dataset, employing Root Mean Square Error (RMSE) as a quantitative metric. Through this evaluation, we aim to demonstrate the network's robustness and its ability to generalize effectively across the dataset, as shown in Fig. \ref{fig:overviewFTGAN}(c), where also the best and worst outputs are presented; the model is not able to capture local features in detail in few instances, particularly due to lack of data. Both ML algorithms were trained using full-wave simulations for data generation. While full wave simulations offer reliability and effectiveness in modeling various scenarios including weakly and highly heterogeneous domains, as well as linear and nonlinear sound wave propagation, they come at a computational expense. Consequently, the data generation process is computationally expensive, leading to a limited dataset. Nonetheless, upon investigating the model's ability to generate \(z_{h}\) layer data for various frequencies, its versatility and adaptability across a spectrum of additional inputs appear promising, as depicted in Fig. \ref{fig:overviewFTGAN}(d).  

\begin{figure*}
    \centering
    \includegraphics[width=1\linewidth]{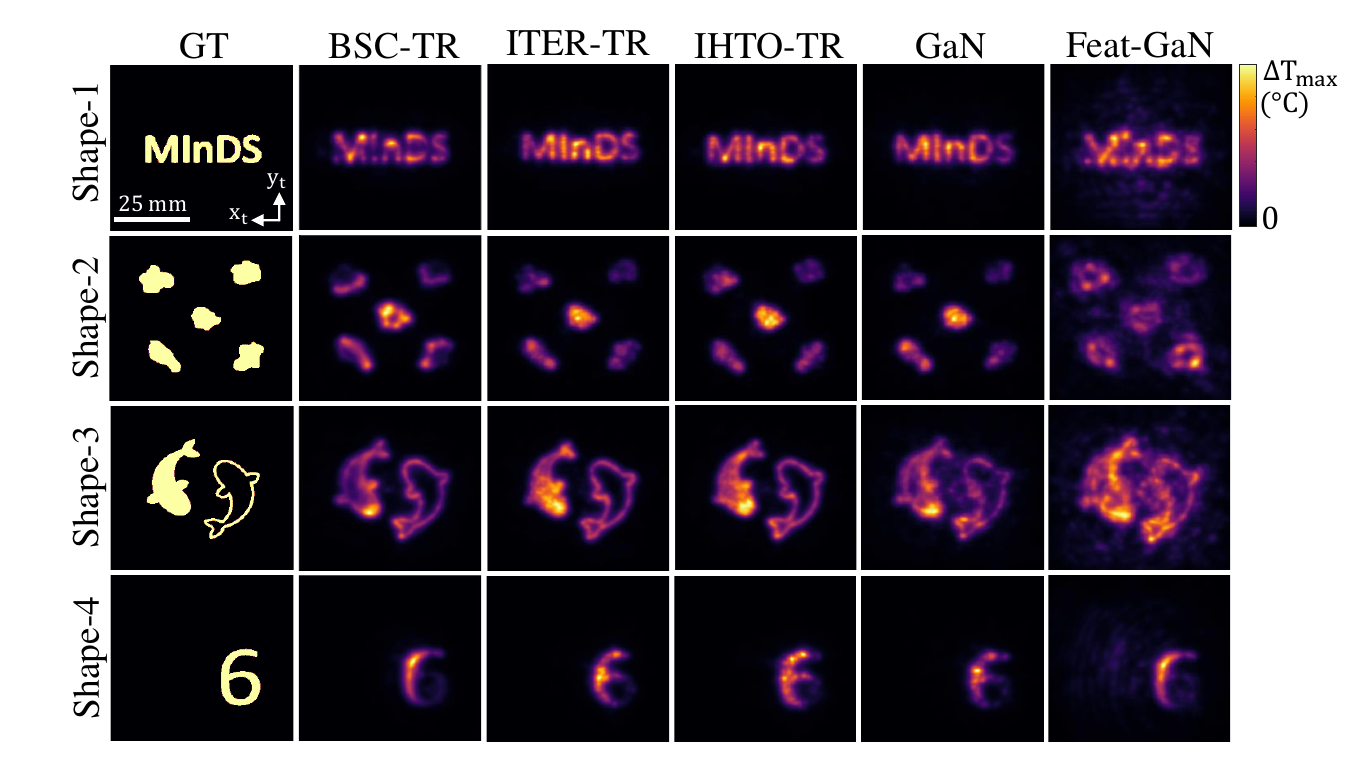}
    \caption{Temperature maps at the target plane generated using different modeling methodologies for AHL. Ground truth (GT) for all four distinct target shapes is provided for comparison.}
    \label{fig:allcompare}
\end{figure*}

\section{Results and Discussions}
\label{ResultsandDisc}

\subsection{Methodological evaluation and comparative analysis}

To ensure a fair comparison of the findings, the normalized temperature distribution at the \(z_t\) is calculated for each modeling approach. Four distinct shape groups, each containing multiple samples are chosen for the imposed target image: (1) Pattern featuring a central focus, as it represents a widely employed type of target image to demonstrate the ability of AHLs to focus complicated patterns. (2)  A multi-focal scenario, where multiple points need to be targeted simultaneously. This case may be particularly relevant for applications where separate targeting is impractical or simultaneous sonication is needed to achieve heating in larger volumes. (3) A complex central focused target with two distinct shapes: empty and filled target shapes. This configuration demonstrates the AHLs' ability to target different types of thermal patterns effectively. (4) Lastly, pattern with a shifted focus is presented, addressing scenarios where the center of the source does not align with the desired focus location for the target. Each of these target shape groups aims to illustrate the potential of the AHL for various cases, facilitating discussions from diverse perspectives on the applied methods. 

Temperature distribution maps for all cases are depicted in Fig. \ref{fig:allcompare}. Upon visually comparing the normalized thermal patterns with the imposed target thermal maps, it's evident that AHLs have the capability to effectively tailor acousto-thermal maps. After closer examination, differences between each method for every target shape become apparent. This suggests that each method may provide unique advantages for various applications, emphasizing the need for a detailed evaluation of the results. Herein, we propose a custom Comprehensive Evaluation Score (CES) (see Eq.\ref{CES}) to incorporate factors such as reconstruction quality, thermal pattern efficiency, computational time, and thermal control capability for extensive comparison of each method. Within the CES value, we also introduce new metrics such as image quality score (IQS) and thermal score (TS). IQS incorporates well-known image quality metrics such as PSNR (Eq. \ref{eq:PSNR}) and SSIM (Eq. \ref{eq:SSIM}), while TS includes thermal efficiency (Eq. \ref{eq:Teff}) and thermal control (TCC) values. The maximum of 1 is indicative of a good performance for the normalized PSNR, SSIM, \(T_{\mathrm{eff}}\) and TCC while smaller values are preferred for the computational time. It's important to note that although there is no universally defined threshold value for PSNR, normalization is essential for calculating the CES. Accordingly, we normalize the PSNR based on the maximum value obtained across all cases. This maximum PSNR value is within the range of values commonly observed in the literature for complex patterning using AHLs \cite{zhong2024real,xu2023programmable,gu2024holographic, lee2022deep, 10.1088/1361-6463/ad5452}. To enhance the applicability and versatility of the proposed CES, adjustments to the equation can also be made to accommodate additional parameters if needed.

\begin{equation}
 CES= IQS + TS - Time
 \label{CES}
\end{equation}

\begin{equation}
   IQS=  1.5 \mid PSNR \mid  + SSIM
    \label{IQM}
\end{equation}

\begin{equation}
   TS=  2 T_{eff}+TCC_{(0,1)}
    \label{TP}
\end{equation}

As briefly discussed in Section \ref{sec:ITER-TR}, the implementation of suitable stopping criteria is essential for iterative modeling approaches for computationally extensive full-wave simulations. An increase in the number of iterations does not necessarily imply an improvement in the quality of the reconstructed thermal image (see \ref{sec:suppiterative}). Even in cases where improvement is observed, it often entails a trade-off between quality and computational time, which may not justify the enhancement. Aiming for the maximum value of quality metrics introduced in Eq. \ref{eq:PSNR}-\ref{eq:Teff} may not serve as a useful termination criterion, as each metric represents different quality criteria, and reaching the maximum value often occurs at different iterations (see \ref{sec:suppiterative}). Therefore, the CES value introduced in this paper not only serves as a valuable comparison metric but also can be utilized as an ending criterion for iterative approaches.

\begin{figure}
    \centering
    \includegraphics[width=1\linewidth]{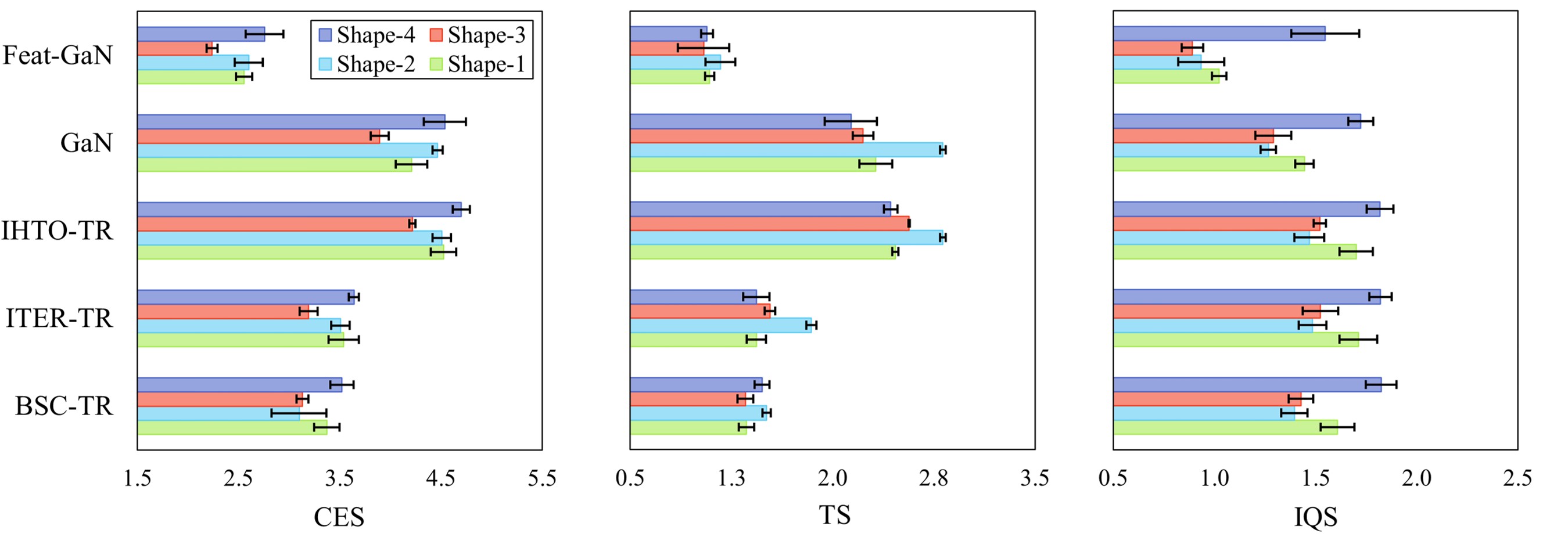}
    \caption{Comprehensive evaluation score (CES), thermal score (TS) and image quiality score (IQS) results for four distinct shape groups each containing four samples and five different methods.}
    \label{fig:lineplot}
\end{figure}

Figure \ref{fig:lineplot} compares the performance metrics for CES, TS, and IQS across all methods. Mean values are determined for four distinct shape groups, each comprising four target shape samples. Standard deviations for these values are calculated using Eq. \ref{stdev}, where n (equal to 4) denotes the number of values per performance metric and shape group, \(v\) and \(\bar{v}\) represents the individual and mean performance metric value, respectively.

\begin{equation}
\label{stdev}
\text{STDEV} = \sqrt{\frac{1}{n-1} \sum_{i=1}^{n} (v_i - \bar{v})^2} 
\end{equation}

As shown in Fig. \ref{fig:lineplot}, IHTO-TR and GaN delivers the highest CES values considering their capability to offer higher thermal pattern efficiency and control capability across different shape groups. BSC-TR and ITER-TR  demonstrates promising performance in terms of IQS, which evaluates image reconstruction efficiency and excludes the thermal effects. Adding iterations to simulations to enhance target pattern results leads to an average 5.9\% improvement in CES value when comparing BSC-TR and ITER-TR, with shape group 2 showing a maximum 13.2\%  improvement. Despite a 21.9\%  increase in TS for shape group 2 with iterative improvements, the CES improvement is less pronounced due to the high computational costs introduced by iterative analysis.

\begin{figure}
    \centering
    \includegraphics[width=0.8\linewidth]{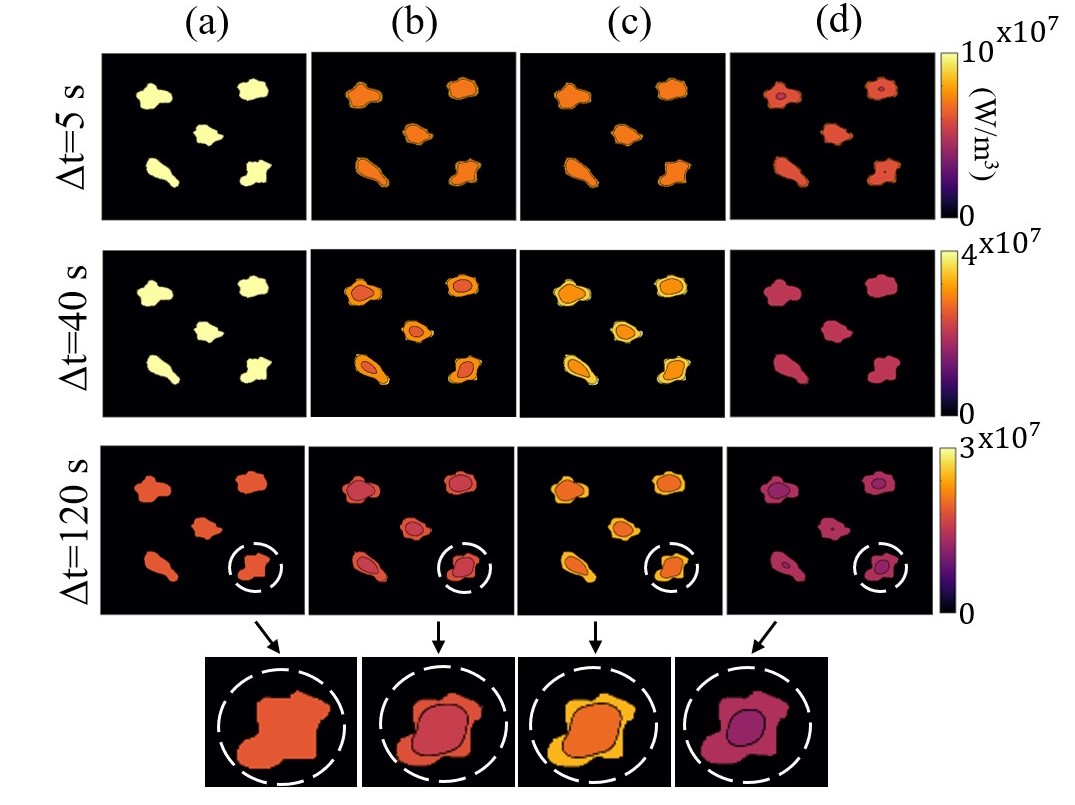}
    \caption{Only thermal analysis heat deposition rate (\(Q_{0}\)) maps are presented to compare the effects of solving the inverse heat transfer optimization (IHTO) problem. (a) Shows the constant heat deposition rate; (b) presents the solution obtained from solving the IHTO; (c) includes the incorporation of blood perfusion rate; and (d) shows the case with cooling time included.}
    \label{fig:cesonlyheat}
\end{figure}

Reflecting on the previous insights given in Section \ref{sec:IHTO-TR} regarding the complicated dynamics of heat transfer, as expected both IHTO-TR and GaN performed well in terms of thermal score. This is primarily attributed to the direct inversion approach employed by IHTO-TR in solving the heat transfer problem and the capability to capture the relation between the \(z_h\) pressure data and the \(z_t\) layer exact temperature values with the GaN algorithm. While the comparison cases depicted in Fig. \ref{fig:allcompare}-\ref{fig:lineplot} are derived from simulations utilizing the BioHeat (Eq.\ref{eq:bioheat}) approximation, in accordance with existing AHL studies, IHTO-TR still achieved higher maximum CES values at lower iteration compared to ITER-TR. For instance, for shape group 4, ITER-TR achieved a CES value of 3.6 by the 4th iteration, whereas IHTO-TR reached 4.7 after just one iteration. To better understand the improvement, additional numerical investigations are done for the thermal part of the multi-physics problem. First, \(Q_t\) is computed by solving the IHTO using Algorithm-1, and the forward problem is run for the optimized \(Q_t\). The computed \(\Delta T_t\) maps are then compared with the forward analysis under the assumption that \(Q_t\) and \(\Delta T_t\) have the same distribution. For this comparison, Eq. \ref{eq:bioheat} is utilized with and without the blood perfusion effect, as well as with integrated cooling time, for varying heating durations. The material properties outlined in Section \ref{Method} are again utilized for the thermal analysis. Additionally,  the blood perfusion effects are quantified
using the parameters \(W_{b} = 0.008330\) 1/s, \(\rho_{b}=1060\) kg/m$^3$ and \(C_{b}=3800\) J/kg.K corresponding to the blood perfusion rate, density, and specific heat, respectively \cite{zhu2009thermal}. The disparity in heat deposition maps is illustrated in Fig. \ref{fig:cesonlyheat} clearly shows the presence of gradients in the heat deposition rate (\(Q_{0}\)) when the IHTP is resolved. At the same time calculated constant \(Q_{0}\) leads to overestimation in values, consequently yielding higher temperatures compared to cases where the IHTP is resolved. Moreover, in real-life applications where cooling periods between sonication are necessary \cite{bredlau2016thermal, coluccia2014first}, solving the IHTP becomes crucial. It not only accurately models heat transfer dynamics but can also be modified to find the most efficient \(Q_{0}\) for specific heating durations and intervals required by the application. Therefore, in instances where precise control over temperature variations is essential, solving the IHTP can facilitate the optimal and effective therapeutic procedure.  Furthermore, in scenarios where the relationship between heat deposition and the pressure field deviates from that described by Eq. \ref{eq:heat deposition}, such as in highly nonlinear regimes, any method neglecting consideration of the thermal field for AHL design expected to lag behind for effective modulation of the focused ultrasound fields. For further insights into the corresponding CES values for the analysis presented in Fig. \ref{fig:cesonlyheat}, readers are referred to \ref{sec:supponlythermal}

From a computational standpoint, approaches based on machine learning, such as GaN and FeatGaN, offer the most significant speed-up compared to other methods. Once successful training is achieved, the system can process a desired thermal map as input and produce the necessary pressure field as output in less than 100 seconds. However, for ITER-TR and IHTO-TR, this processing time increases to over 100,000 seconds with iterations. Upon analysis of Fig. \ref{fig:allcompare} and Fig. \ref{fig:lineplot}, it becomes evident that the performance of the case-specific GaN technique surpasses that of the more complex Feat-GaN model, primarily because the problem itself is configured to account for frequency-dependent variations. Furthermore, the datasets utilized for training ML-based approaches are constrained by computationally expensive but reliable full wave simulations. As mentioned earlier, the roadmaps outlined in this study are adaptable to different propagation models, and there is potential to integrate faster methods into the data generation process to facilitate the creation of larger datasets and enhance accuracy. The Feat-GaN model, in particular, shows promise as a future AHL design methodology, offering a more versatile platform capable of accommodating larger datasets.

\subsection{Target pattern complexity}

\begin{figure}
    \centering
    \includegraphics[width=1\linewidth]{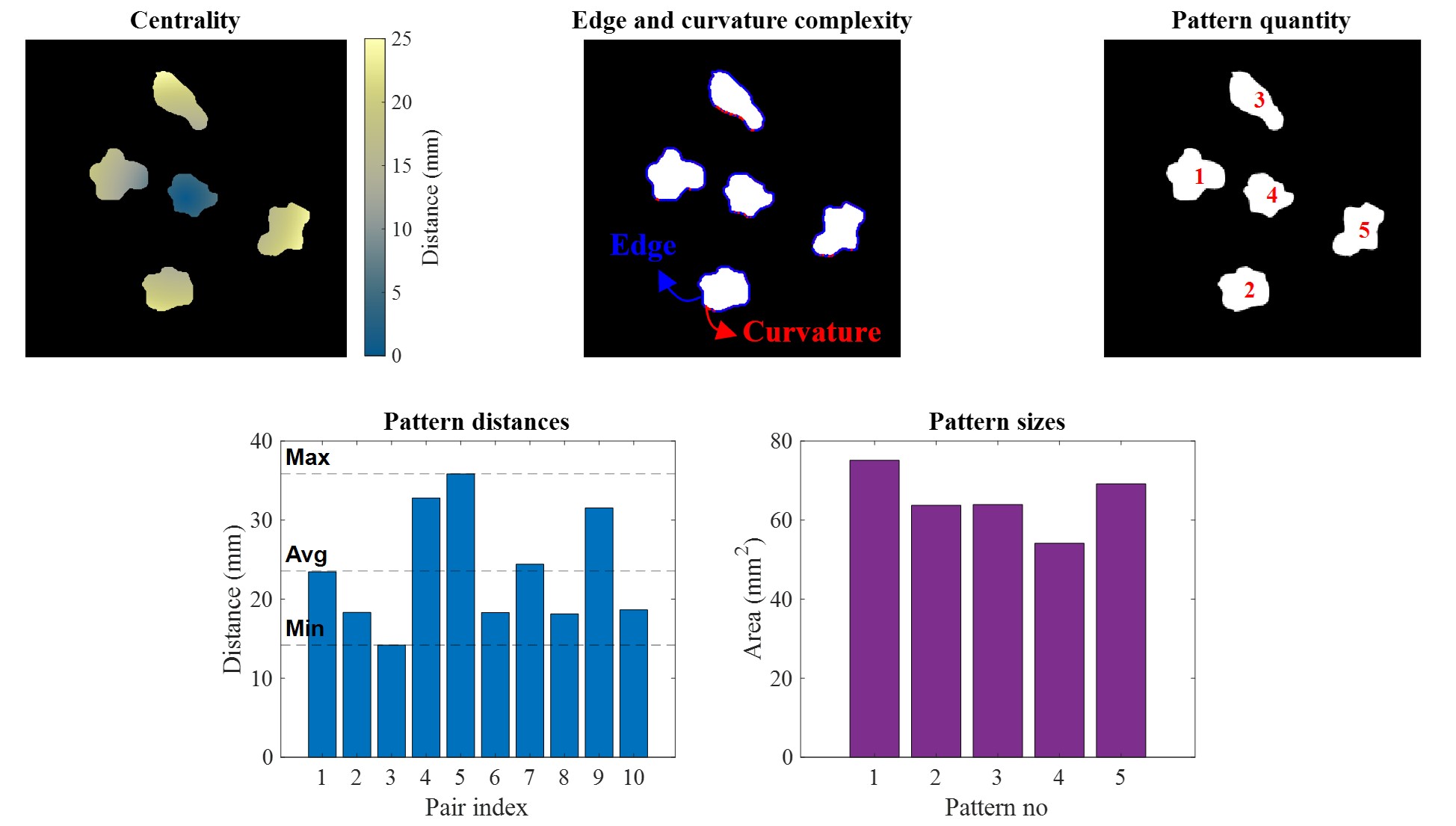}
    \caption{Target pattern complexity assesment based on centrality, edge and curve complexity, pattern quantity, pattern distances and pattern sizes.}
    \label{fig:complexity}
\end{figure}

Finally, as shown in thermal maps in Fig.\ref{fig:allcompare} and the comparisons in Fig.\ref{fig:lineplot},  CES, TS, and IQS values vary based on the chosen modeling method and shape group. Considering that each shape group exhibits distinct characteristics, a deeper analysis of their pattern complexities can enhance our understanding of the impact on targeting efficiency. There are several ways to differentiate imposed target patterns. Shape complexity can be assessed through human perception, with more edges and curves generally indicating higher complexity \cite{bazazian2022perceptually} or mathematical measures such as curvature entropy \cite{page2003shape} and shape randomness \cite{chen2005estimating} can be used to quantify shape complexity according to the specific demands of an application. Given the clinical focus of AHL-assisted thermal patterning in this study, the following metrics are employed to quantify the complexity of the shape groups analyzed: (1) Centrality, (2) Edge and curve complexity, (3) Pattern quantity, (4) Pattern distances and (5) Pattern sizes. Centrality is quantified by calculating the Euclidean distance from the center to each pattern within the shape group. The x- and y-coordinates of the center are determined, and the distances from these coordinates to each target pixel are computed. The minimum, maximum, and average distances are then used to evaluate the overall centrality of each target pattern within the shape group. Edge complexity is assessed by applying the Sobel filter to detect all edges and curvature complexity is quantified by calculating the change in angles along the pattern boundaries. Pattern quantity is determined by identifying connected components within the target image, with each distinct pattern labeled and counted. The areas of these patterns are measured in pixels, and the mean area is calculated to assess the overall size distribution within the shape group. Additionally, for images containing multiple patterns, the pairwise Euclidean distances between the centroids of the patterns are computed, with the maximum, minimum, and average pattern distances. Each pair index represents a specific combination of pattern centroids, allowing for a detailed assessment of the spatial relationships. An example of target pattern complexity analysis for a multi focal pattern from shape group 2 is illustrated in Fig. \ref{fig:complexity} as a reference. A summary of the pattern complexity investigations for differentiating each shape group based on their characteristics is provided in Table \ref{tab:shape_summary}.

\begin{table}[t]
\centering
\caption{Summary of pattern complexity characteristics for each shape group}
\label{tab:shape_summary}
\resizebox{\textwidth}{!}{
\begin{tabular}{ccccc} 
\hline
\hline
\textbf{} & \textbf{Shape-1} & \textbf{Shape-2} & \textbf{Shape-3} & \textbf{Shape-4} \\ 
\midrule
\textbf{Centrality} & Centralized & Centralized & Centralized & Decentralized \\
\textbf{Edge and curve complexity} & Medium & Medium & High & Low \\
\textbf{Pattern quantity} & High & High & Low & Low \\  \textbf{Pattern distances} & Clustered & Dispersed & Dispersed & Dispersed \\
\textbf{Pattern sizes} &Small &Medium &Large &Large\\
\hline
\hline

\end{tabular}
}
\end{table}

Combined investigations of Fig.\ref{fig:lineplot} and Table.\ref{tab:shape_summary} have shown that the lowest TS value is obtained for shape group 4, where the target pattern is located outside made it hard to achieve an full focus outside of the transducer's high amplitude region. For effective targetting, iterative or optimization-based
algorithms are expected to perform better than non-iterative approaches, as
they allow the user to enforce the desired parameters to reach maximum effi-
ciency. The lowest CES is observed for shape group 3, which exhibits high edge and curve complexity compared to the other shape groups. However, despite the close proximity of target patterns to the center, their larger size and dispersed distribution still result in relatively high TS values compared to shape groups with smaller, clustered patterns or those located further from the center. Shape groups 1 and 2 show similar CES values due to their centralized locations and comparable edge and curve complexities, as well as pattern quantities. However, shape group 2 yields significantly higher TS values due to its dispersed pattern distribution and larger pattern sizes, which make targeting easier. The pattern complexity investigations presented in this section offer a way to compare and distinguish between shape groups, allowing for informed decisions tailored to specific applications and target patterns. Depending on the application, different metrics can be used to assess pattern complexity. Threshold values can be established to guide the adoption of different methods based on whether the complexity exceeds or falls below certain levels. For more detail on how the Table \ref{tab:shape_summary} is formed, readers can refer to the additional data provided in \ref{supppatterncomplexity}.

\subsection{Clinical applications of holographic focused ultrasound}

Focused ultrasound is a versatile and rapidly evolving technology with significant potential to revolutionize medical treatments across various fields. Its non-invasive nature and ability to precisely target tissues make it a promising tool for clinical applications and emerging therapeutic strategies. Current uses include thermal ablation, drug delivery, neuromodulation \cite{legon2014transcranial, legon2022noninvasive, strohman2024low} and immunomodulation with emerging applications including histotripsy \cite{xu2021histotripsy}, sonodynamic therapy, gene delivery and blood-brain barrier (BBB) opening \cite{meng2021applications, rabut2020ultrasound, izadifar2020introduction}. In all of these applications, precise application of the ultrasound field to the target tissue volume is critical to avoid unwanted effects in adjacent, overlying, or surrounding tissue. Focused ultrasound is currently FDA-approved to treat uterine fibroids, musculoskeletal indications including bone metastases and osteoid osteomas, urological indications including benign prostatic hyperplasia and prostate cancer, and neurological indications including essential tremor, Parkinson’s dyskinesia, and tremor. Several additional neurological applications are in the clinical trial phase including depression, post-traumatic stress disorder, brain metastases, bipolar disorder, cerebral palsy, neuropathic pain, and obsessive-compulsive disorder \cite{FUS}.

In clinical applications, the temperature increase caused by ultrasound necessitates a carefully designed therapeutic computational modeling strategy that considers acousto-thermal properties such as intensity, duration, and duty cycle, along with local tissue characteristics like absorption, thermal diffusion, and perfusion. The objective is to ensure high heating rates within the focal target region while reducing heat exposure to the skin, mucosa, and subcutaneous tissues \cite{ebbini2015ultrasound}. Therefore, computational approaches that consider local tissue properties and acousto-thermal interactions, such as IHTO-TR and GaN methods proposed in this study, can be highly valuable. Drawing from current FUS applications, AHL-assisted FUS holds promise for future expansion into treating a variety of clinical conditions similar to those currently addressed by traditional FUS methods, including prostate, breast, liver, renal, and bone tumors, as well as pancreatic cancer and uterine fibroids \cite{zhou2011high}. In applications of mild hyperthermia for optimal drug delivery, key performance criteria include temperature accuracy, temporal control, homogeneity of heating, and conformal heating of the desired region \cite{partanen2012mild}. These criteria align with the discussions in this paper regarding various methodologies, where AHLs can offer significant improvements to the aforementioned points when a suitable modeling approach is employed. For transcranial FUS,  neurological indications are particularly challenging due to the skull that highly attenuates and aberrates ultrasound and also the morphology of different brain structures that can have highly irregular shapes and volumes that are also in close proximity to adjacent neuronal structures and brain circuitry that necessitate precise shaping and steering. When properly trained, ML-based approaches such as GaN and Feat-GaN can make a significant difference by efficiently accounting for the high attenuation and aberrations, which typically require computationally expensive acoustic wave propagation models. Moreover, ML-based techniques offer the advantage of being trainable for specific treatments based on parameters such as thermal dose and spot-size, which are often linked to efficient treatment processes in thermal neuromodulation applications \cite{sammartino2021thermal}.

\section{Conclusion}

This paper presents a thorough investigation of computational acousto-thermal modeling approaches for AHL design. These methodologies are divided into three types: (1) pressure-based approaches (BSC-TR and ITER-TR), which focus on achieving a precise pressure field; (2) thermal-based methods (IHTO-TR), which incorporate inverse heat transfer optimization into acoustic-wave propagation simulations; and (3) machine learning-based techniques (GaN and Feat-GaN), which use specific and generic algorithms for direct inversion of the thermal field into holographic representations. For comparison, we examined four distinct target shape groups, each containing multiple samples, to account for the unique characteristics across diverse scenarios and thereby broaden the applicability of the methodologies outlined in this paper. To facilitate a comprehensive assessment, we introduced a novel comparison metric encompassing image quality, thermal efficiency and control, and computational time parameters. Our results demonstrate that high CES values between 4.2 and 4.7 can be obtained for each shape group by employing the appropriate modeling methodology. The importance of integrating heat transfer dynamics into the generation of thickness maps for 3D production of AHLs is highlighted in this paper, emphasizing how the quality of the thermal patterns and computational efficiency improve when utilizing IHTO-TR or GaN. Furthermore, heat transfer analysis alone showed that this integration facilitates the determination of optimum heat deposition rates tailored to the specific requirements of the intended application. Especially for future real-life applications requiring consideration of biological features, non-uniform characteristics, and non-Fourier behaviors, temperature and machine learning-based approaches emerge as promising options compared to pressure-based inverse modeling approaches. GaN and Feat-GaN methods enabled rapid AHL design generation in less than 100 seconds, suitable for 3D manufacturing and subsequent use. Additionally, they offer the advantage of real-time pattern adjustments, allowing simultaneous alteration of AHL designs without the need to rerun acousto-thermal simulations for each variation.

In summary, the BSC-TR method is ideal for achieving fast solutions in cases where the medium is lightly heterogeneous and the target pattern has a geometric shape that is centralized, not complex, with low pattern number and dispersed distribution. ITER-TR is better suited for cases where the imposed target pattern includes more intricate geometric details, such as curves, varying angles, edges, and holes, or when the target shape is not aligned with the center of focus, while the medium remains lightly heterogeneous. When heat transfer effects are significant, the medium is highly heterogeneous, and precise optimization of heat deposition is needed, IHTO-TR is recommended regardless of the shape's complexity. GaN is preferred for quick solutions, especially when computational time is critical for the intended application. GaN also accounts for acousto-thermal effects, making it suitable for cases where the domain is highly heterogeneous. It is ideal for applications where parameters such as frequency, thermophysical and acoustical properties, and distances remain constant, but the imposed pattern is subject to change. Feat-GaN, on the other hand, is efficient for cases with moderate heat transfer effects and shapes with lower structural complexity. This algorithm can be trained to accommodate variations in model parameters, making it suitable for more generic applications. In addition to the methodology guidelines provided in the paper, Table \ref{tab:comprehensive} offers a concise summary of key comparison points for reference.

\begin{table}[h]
   \centering     
    \caption{Summary of key comparison points for different modeling methodologies for the design of acoustic holographic lenses (AHLs) for acousto-thermal manipulation}
    \resizebox{\textwidth}{!}{
    \begin{tabular}{lcccccc}
    \midrule
    \midrule
          \textbf{} & \textbf{ BSC-TR} & \textbf{ ITER-TR} & \textbf{IHTO-TR} & \textbf{GaN} & \textbf{Feat-GaN} \\
          \midrule
          \textbf{Inverted field} & Pressure & Pressure & Thermal & Thermal & Normalized thermal \\
          \textbf{Computational speed} & Moderate & Slow & Moderate & Fast & Fast \\
          \textbf{Thermal gradient control} & No & No & Yes & Yes & Limited \\
          \textbf{Implementation difficulty} & Easy & Moderate & Moderate & Easy & Easy \\
          \textbf{Dynamic pattern adjustment} & No & No & No & Yes & Yes \\
          \textbf{Scalability} & Limited & No & No & Yes & Yes \\
          \textbf{Immediate parameter adaptation} & No & No & No & No & Yes \\
          \midrule
          \midrule
    \end{tabular}
    }
    \label{tab:comprehensive}
\end{table}

\appendix

\section{Impact of amplitude parameter (\textit{m} on phase-only hologram calculations}
\label{berd}
\begin{figure}[H]
    \centering
    \includegraphics[width=0.7\linewidth]{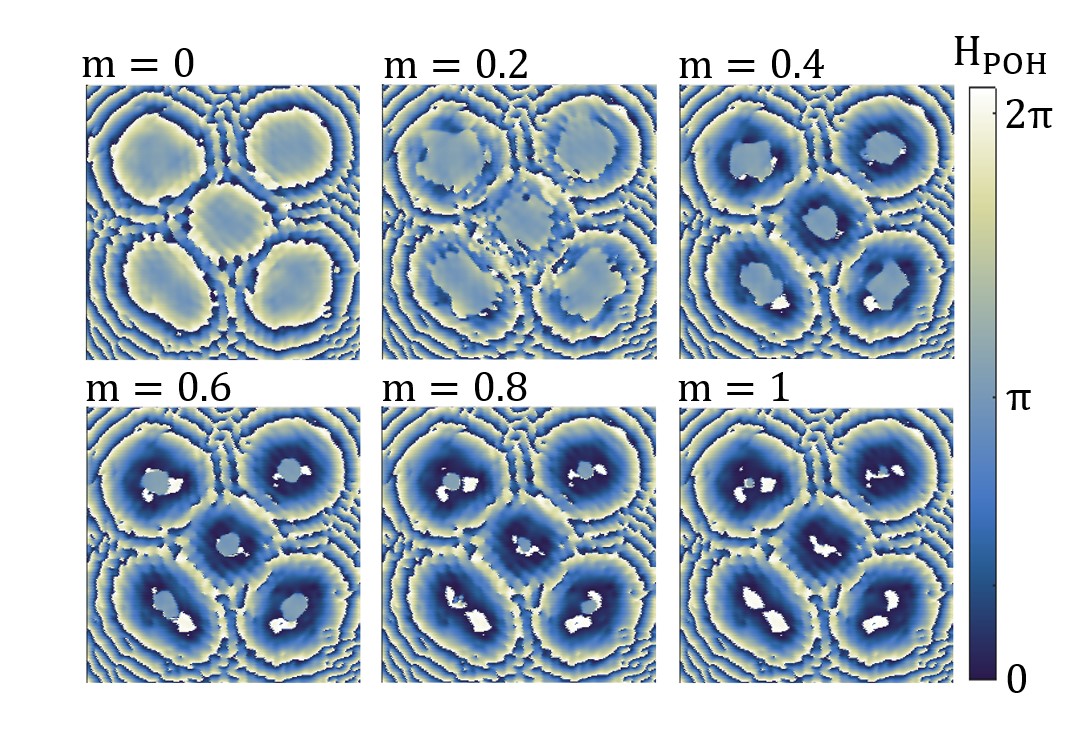}
    \caption{Phase only hologram (\(H_{POH}\)) change with increasing value of amplitude parameter (\(m\)).}
    \label{fig:BERDm}
\end{figure}

Figure \ref{fig:BERDm} demonstrates the impact of adjusting the amplitude parameter (m) on the phase-only hologram (\(H_{POH})\) calculations using the BSC-TR method. The series of phase maps in Fig. \ref{fig:BERDm} reveal noticeable variations corresponding to different values of m. By manipulating m, the holographic reconstruction process can be fine-tuned, enhancing the pattern efficiency when BSC-TR method is employed for modeling.

\section{Thermal pattern variations with increasing ITER-TR iterations}
\label{sec:suppiterative}

\begin{figure}[H]
    \centering
    \includegraphics[width=0.9\linewidth]{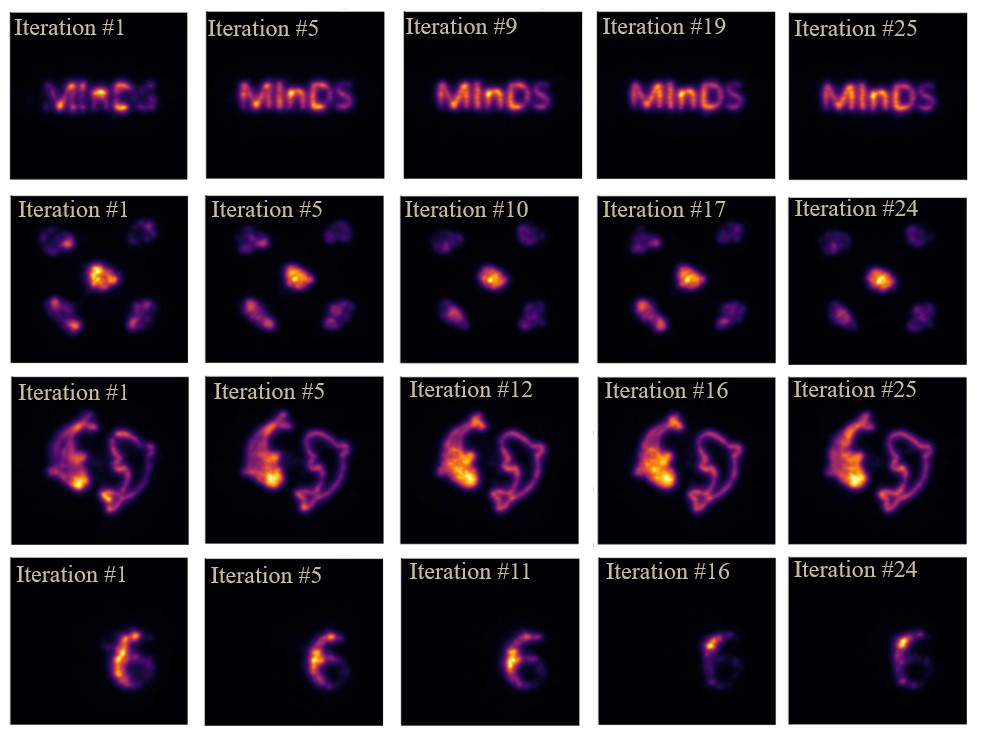}
    \caption{Change of thermal maps with iterations for four different target shapes.}
    \label{fig:suppiter}
\end{figure}

\begin{figure}[H]
    \centering
    \includegraphics[width=0.7\linewidth]{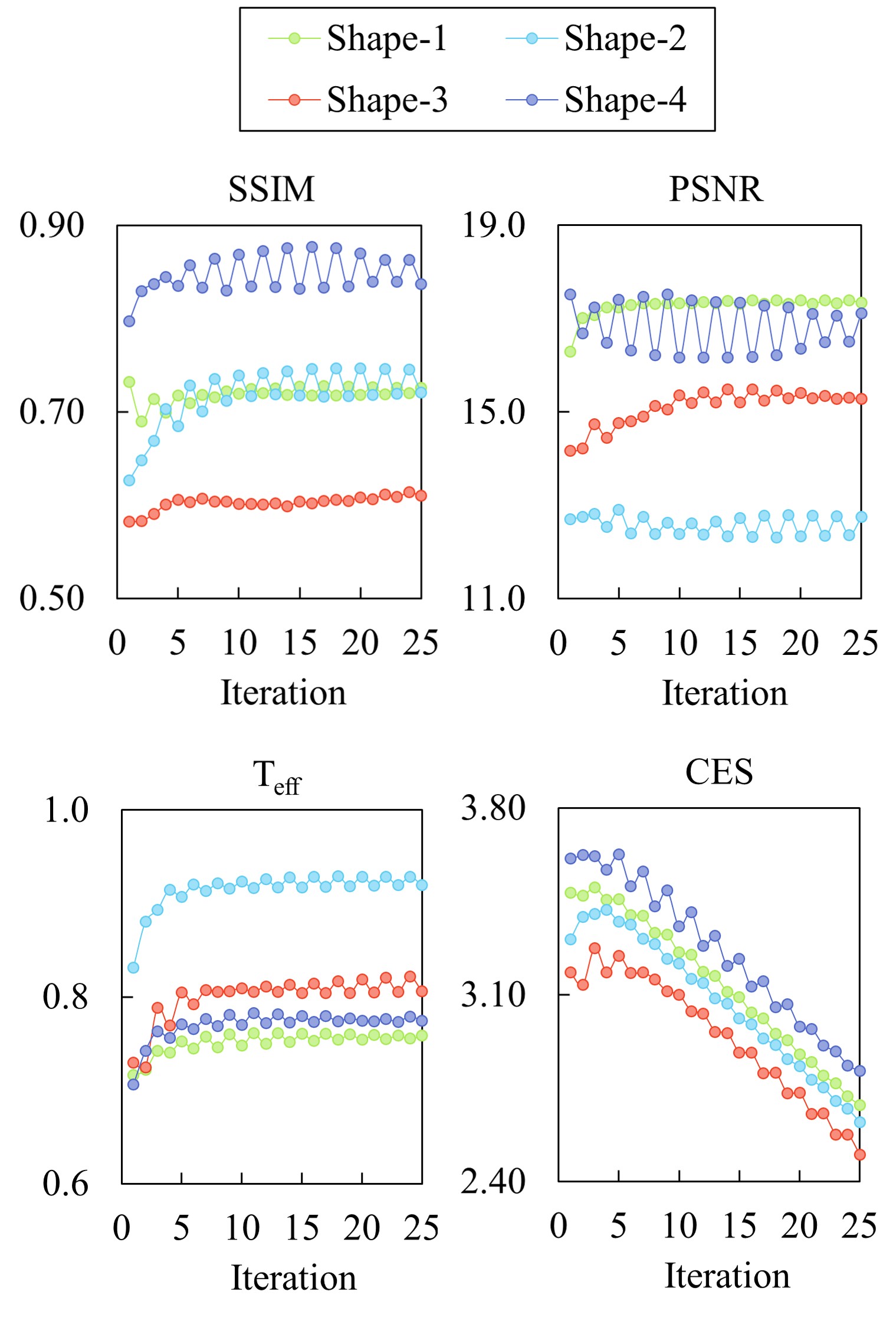}
    \caption{Dynamic variation of the structural similarity index (SSIM), peak-signal-to-noise ratio (PSNR), thermal efficiency (\(T_{eff}\)) and comprehensive evaluation score (CES) with increasing iterations in ITER-TR method.}
    \label{fig:suppiter2}
\end{figure}

Figure \ref{fig:suppiter} provides a reference for the variation of the thermal map with increasing iterations. Specific iterations are selected to highlight these changes effectively. Noticeable improvements in pattern shape are evident for shape groups 1 and 3 as the number of iterations increases; however, after a certain point, these changes become minimal. Conversely, the improvements in shape groups 2 and 4 are less pronounced.  For a better comparison, variations in error metrics should also be investigated (see Fig. \ref{fig:suppiter2}).

\section{Heat transfer analysis on the impact of inverse problem solutions on computed heat deposition rates}
\label{sec:supponlythermal}

\begin{figure}[H]
    \centering
    \includegraphics[width=0.45\linewidth]{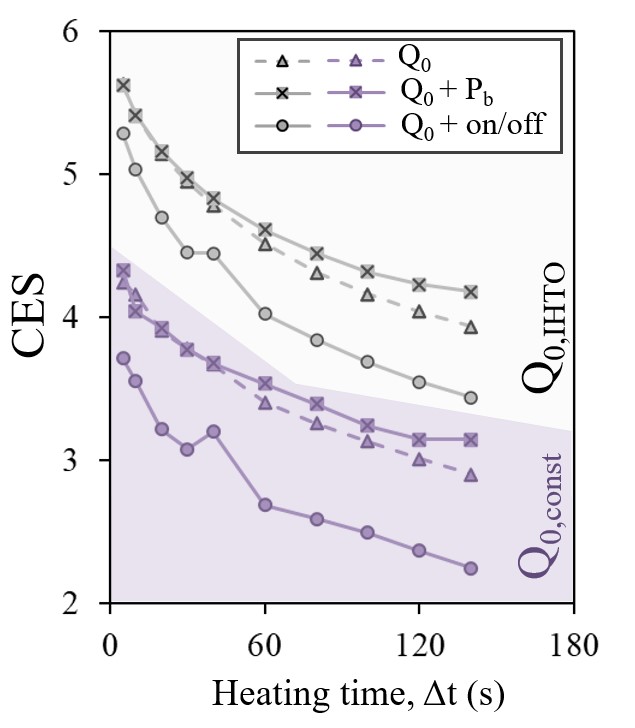}
    \caption{Comperative analysis of Comprehensive Evaluation Score (CES) values using different heat deposition rates}
    \label{fig:onlythermalces}
\end{figure}

Figure \ref{fig:onlythermalces} illustrates the variations in Comprehensive Evaluation Score (CES) values calculated using different heat deposition rates. \( Q_{0,const}\) represents a constant uniform rate, while  \( Q_{0,IHTO}\) corresponds to the rate derived from inverse heat transfer optimization (IHTO). Three scenarios are examined: (1) The baseline case (\( Q_{0}\)), as detailed in the main manuscript for method comparison. (2) \( Q_{0}+P_{b}\), which includes blood perfusion rates. (3) \( Q_{0}+on/off\), involving a heating phase followed by a cooling period to manage temperatures post-ultrasound absorption. The CES variations are plotted against heating time, highlighting the importance of inverse heat transfer problem solutions in optimizing acoustic holographic lens design and supporting the results presented in Fig. \ref{fig:cesonlyheat}

\section{Supplementary data for target pattern complexity analysis}
\label{supppatterncomplexity}

\begin{figure}[H]
    \centering
    \includegraphics[width=0.8\linewidth]{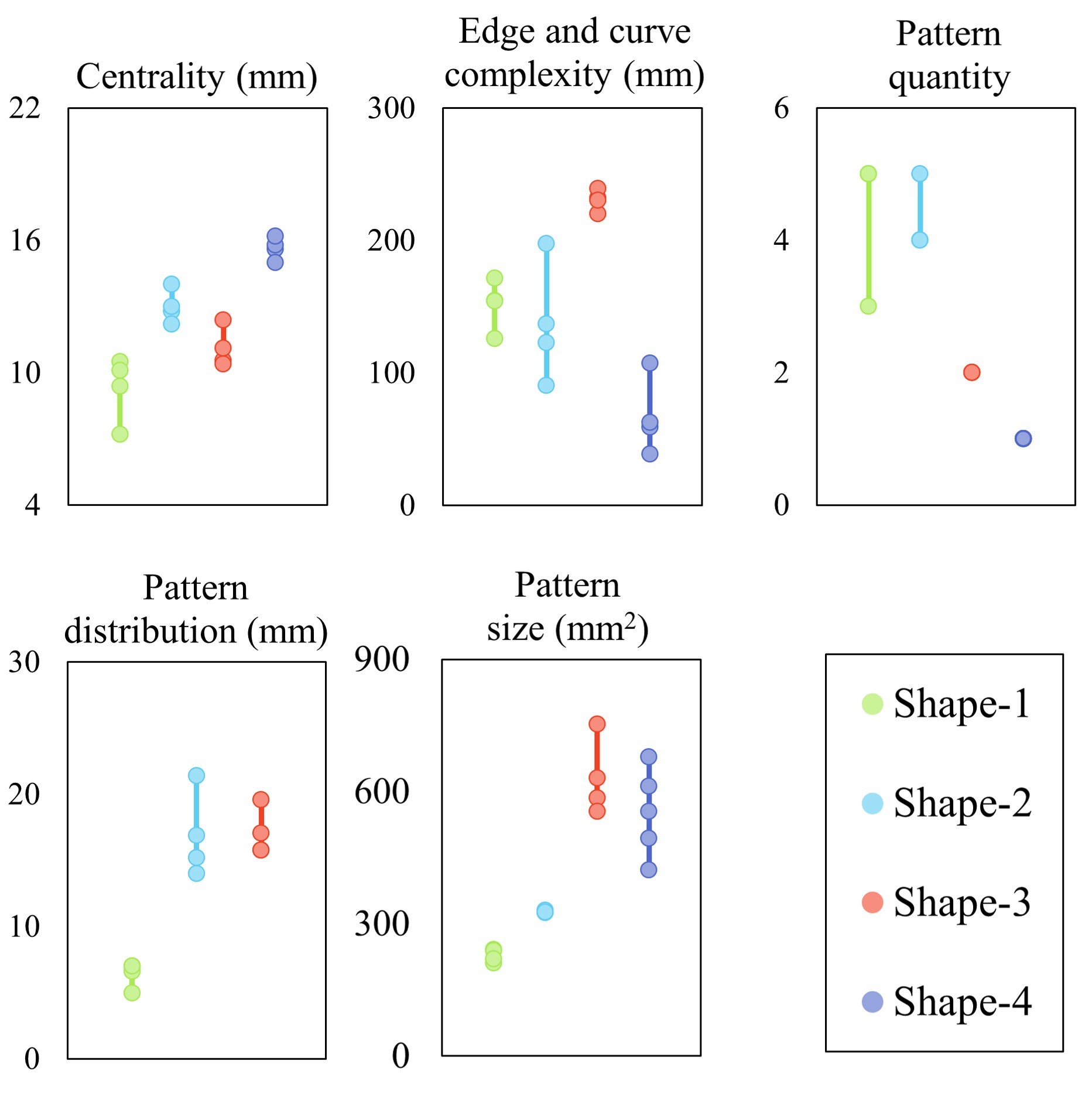}
    \caption{Categorization of shape groups based on centrality, edge and curve complexity, pattern quantity, distribution, and size to evaluate target pattern complexity. }
    \label{fig:patterncomplex}
\end{figure}

This figure provides a detailed visualization of the complexity characteristics of target patterns across different shape groups, each containing four distinct target patterns. The metrics used—centrality, edge and curve complexity, pattern quantity, pattern distribution, and pattern size—are selected to comprehensively evaluate the complexity of these target patterns. The results presented here offer a detailed version of the target pattern complexity summary found in Table \ref{tab:shape_summary} of the main manuscript.

\bibliographystyle{elsarticle-num} 
\bibliography{refs}

\begin{thebibliography}{10}
\expandafter\ifx\csname url\endcsname\relax
  \def\url#1{\texttt{#1}}\fi
\expandafter\ifx\csname urlprefix\endcsname\relax\def\urlprefix{URL }\fi
\expandafter\ifx\csname href\endcsname\relax
  \def\href#1#2{#2} \def\path#1{#1}\fi

\bibitem{ku2001applications}
H.~Ku, F.~Siu, E.~Siores, J.~Ball, A.~Blicblau, Applications of fixed and variable frequency microwave (vfm) facilities in polymeric materials processing and joining, Journal of Materials Processing Technology 113~(1-3) (2001) 184--188.

\bibitem{zohm2000thermal}
H.~Zohm, E.~Kasper, P.~Mehringer, G.~M{\"u}ller, Thermal processing of silicon wafers with microwave co-heating, Microelectronic engineering 54~(3-4) (2000) 247--253.

\bibitem{ibarra2009comparative}
C.~Ibarra-Castanedo, J.-M. Piau, S.~Guilbert, N.~P. Avdelidis, M.~Genest, A.~Bendada, X.~P. Maldague, Comparative study of active thermography techniques for the nondestructive evaluation of honeycomb structures, Research in Nondestructive Evaluation 20~(1) (2009) 1--31.

\bibitem{bhargava2019coupling}
A.~Bhargava, S.~Shahab, Coupling of nonlinear shape memory polymer cantilever dynamics with focused ultrasound field, Smart Materials and Structures 28~(5) (2019) 055002.

\bibitem{baghban2017estimation}
M.~Baghban, M.~B. Ayani, Estimation of surface heat flux in a one-dimensional hyperbolic bio-heat conduction problem with temperature-dependent properties during thermal therapy, Journal of the Brazilian Society of Mechanical Sciences and Engineering 39 (2017) 1479--1489.

\bibitem{escoffre2015therapeutic}
J.-M. Escoffre, A.~Bouakaz, Therapeutic ultrasound, Vol. 880, Springer, 2015.

\bibitem{zhu2024shape}
Y.~Zhu, K.~Deng, J.~Zhou, C.~Lai, Z.~Ma, H.~Zhang, J.~Pan, L.~Shen, M.~D. Bucknor, E.~Ozhinsky, et~al., Shape-recovery of implanted shape-memory devices remotely triggered via image-guided ultrasound heating, Nature Communications 15~(1) (2024) 1123.

\bibitem{xu2024safety}
R.~Xu, B.~E. Treeby, E.~Martin, Safety review of therapeutic ultrasound for spinal cord neuromodulation and blood--spinal cord barrier opening, Ultrasound in Medicine \& Biology (2024).

\bibitem{ter2007high}
G.~ter Haar, C.~Coussios, High intensity focused ultrasound: physical principles and devices, International journal of hyperthermia 23~(2) (2007) 89--104.

\bibitem{melde2016holograms}
K.~Melde, A.~G. Mark, T.~Qiu, P.~Fischer, Holograms for acoustics, Nature 537~(7621) (2016) 518--522.

\bibitem{brown2020stackable}
M.~D. Brown, B.~T. Cox, B.~E. Treeby, Stackable acoustic holograms, Applied Physics Letters 116~(26) (2020).

\bibitem{zhang2020acoustic}
J.~Zhang, Y.~Tian, Y.~Cheng, X.~Liu, Acoustic holography using composite metasurfaces, Applied Physics Letters 116~(3) (2020).

\bibitem{jimenez2019holograms}
S.~Jim{\'e}nez-Gamb{\'\i}n, N.~Jim{\'e}nez, J.~M. Benlloch, F.~Camarena, Holograms to focus arbitrary ultrasonic fields through the skull, Physical Review Applied 12~(1) (2019) 014016.

\bibitem{sallam2021holographic}
A.~Sallam, V.~C. Meesala, M.~R. Hajj, S.~Shahab, Holographic mirrors for spatial ultrasound modulation in contactless acoustic energy transfer systems, Applied Physics Letters 119~(14) (2021).

\bibitem{bakhtiari2018acoustic}
M.~Bakhtiari-Nejad, A.~Elnahhas, M.~R. Hajj, S.~Shahab, Acoustic holograms in contactless ultrasonic power transfer systems: Modeling and experiment, Journal of Applied Physics 124~(24) (2018).

\bibitem{andres2022thermal}
D.~Andr{\'e}s, J.~Vappou, N.~Jim{\'e}nez, F.~Camarena, Thermal holographic patterns for ultrasound hyperthermia, Applied Physics Letters 120~(8) (2022).

\bibitem{andres2023holographic}
D.~Andr{\'e}s, I.~Rivens, P.~Mouratidis, N.~Jim{\'e}nez, F.~Camarena, G.~Ter~Haar, Holographic focused ultrasound hyperthermia system for uniform simultaneous thermal exposure of multiple tumor spheroids, Cancers 15~(9) (2023) 2540.

\bibitem{li2022generating}
T.~Li, J.~Li, L.~Bo, Z.~Tian, Generating multi-pixel thermal images through an acousto-thermal effect, in: ASME International Mechanical Engineering Congress and Exposition, Vol. 86670, American Society of Mechanical Engineers, 2022, p. V005T07A010.

\bibitem{kim2023nanoparticle}
J.~Kim, S.~Kasoji, P.~G. Durham, P.~A. Dayton, Nanoparticle-epoxy composite molding for undeformed acoustic holograms with tailored acoustic properties, IEEE Transactions on Ultrasonics, Ferroelectrics, and Frequency Control (2023).

\bibitem{gu2015modeling}
J.~Gu, Y.~Jing, Modeling of wave propagation for medical ultrasound: a review, IEEE transactions on ultrasonics, ferroelectrics, and frequency control 62~(11) (2015) 1979--1992.

\bibitem{fushimi2021acoustic}
T.~Fushimi, K.~Yamamoto, Y.~Ochiai, Acoustic hologram optimisation using automatic differentiation, Scientific reports 11~(1) (2021) 12678.

\bibitem{sallam2023nonlinear}
A.~Sallam, S.~Shahab, Nonlinear acoustic holography with adaptive sampling, IEEE Transactions on Ultrasonics, Ferroelectrics, and Frequency Control (2023).

\bibitem{ferri2019enhanced}
M.~Ferri, J.~M. Bravo, J.~Redondo, J.~V. S{\'a}nchez-P{\'e}rez, Enhanced numerical method for the design of 3-d-printed holographic acoustic lenses for aberration correction of single-element transcranial focused ultrasound, Ultrasound in medicine \& biology 45~(3) (2019) 867--884.

\bibitem{jimenez2021acoustic}
S.~Jim{\'e}nez-Gamb{\'\i}n, N.~Jim{\'e}nez, A.~N. Pouliopoulos, J.~M. Benlloch, E.~E. Konofagou, F.~Camarena, Acoustic holograms for bilateral blood-brain barrier opening in a mouse model, IEEE Transactions on Biomedical Engineering 69~(4) (2021) 1359--1368.

\bibitem{sallam2024gradient}
A.~Sallam, C.~Cengiz, M.~Pewekar, E.~Hoffmann, W.~Legon, E.~Vlaisavljevich, S.~Shahab, Gradient descent optimization of acoustic holograms for transcranial focused ultrasound, arXiv preprint arXiv:2401.14756 (2024).

\bibitem{fink1992}
M.~Fink, Time reversal of ultrasonic fields. i. basic principles, IEEE Transactions on Ultrasonics, Ferroelectrics, and Frequency Control 39~(5) (1992) 555--566.
\newblock \href {https://doi.org/10.1109/58.156174} {\path{doi:10.1109/58.156174}}.

\bibitem{kinsler2000fundamental}
L.~E. Kinsler, A.~R. Frey, A.~B. Coppens, J.~V. Sanders, Fundamentals of acoustics, John wiley \& sons, 2000.

\bibitem{fink1999time}
M.~Fink, Time-reversed acoustics, Scientific American 281~(5) (1999) 91--97.

\bibitem{treeby2013acoustic}
B.~E. Treeby, Acoustic attenuation compensation in photoacoustic tomography using time-variant filtering, Journal of biomedical optics 18~(3) (2013) 036008--036008.

\bibitem{sallam2022nonlinear}
A.~Sallam, S.~Shahab, On nonlinear effects in holographic-modulated ultrasound, Applied Physics Letters 121~(20) (2022).

\bibitem{filonenko2001effect}
E.~Filonenko, V.~Khokhlova, Effect of acoustic nonlinearity on heating of biological tissue by high-intensity focused ultrasound, Acoustical Physics 47 (2001) 468--475.

\bibitem{parker2022power}
K.~Parker, Power laws prevail in ultrasound-tissue interactions, Physics in medicine and biology 67~(9) (2022).

\bibitem{pennes1948analysis}
H.~H. Pennes, Analysis of tissue and arterial blood temperatures in the resting human forearm, Journal of applied physiology 1~(2) (1948) 93--122.

\bibitem{chen1980microvascular}
M.~M. Chen, K.~R. Holmes, Microvascular contributions in tissue heat transfer, Annals of the New York Academy of Sciences 335~(1) (1980) 137--150.

\bibitem{yang2007expanding}
D.~Yang, M.~C. Converse, D.~M. Mahvi, J.~G. Webster, Expanding the bioheat equation to include tissue internal water evaporation during heating, IEEE Transactions on Biomedical Engineering 54~(8) (2007) 1382--1388.

\bibitem{yuan2008numerical}
P.~Yuan, Numerical analysis of temperature and thermal dose response of biological tissues to thermal non-equilibrium during hyperthermia therapy, Medical engineering \& physics 30~(2) (2008) 135--143.

\bibitem{LIU20101138}
K.-C. Liu, H.-T. Chen, \href{https://www.sciencedirect.com/science/article/pii/S129007291000058X}{Investigation for the dual phase lag behavior of bio-heat transfer}, International Journal of Thermal Sciences 49~(7) (2010) 1138--1146.
\newblock \href {https://doi.org/https://doi.org/10.1016/j.ijthermalsci.2010.02.007} {\path{doi:https://doi.org/10.1016/j.ijthermalsci.2010.02.007}}.
\newline\urlprefix\url{https://www.sciencedirect.com/science/article/pii/S129007291000058X}

\bibitem{gupta2019non}
P.~Gupta, A.~Srivastava, Non-fourier transient thermal analysis of biological tissue phantoms subjected to high intensity focused ultrasound, International Journal of Heat and Mass Transfer 136 (2019) 1052--1063.

\bibitem{andreozzi2019modeling}
A.~Andreozzi, L.~Brunese, M.~Iasiello, C.~Tucci, G.~P. Vanoli, Modeling heat transfer in tumors: A review of thermal therapies, Annals of biomedical engineering 47 (2019) 676--693.

\bibitem{singh2020thermal}
S.~Singh, R.~Melnik, Thermal ablation of biological tissues in disease treatment: A review of computational models and future directions, Electromagnetic biology and medicine 39~(2) (2020) 49--88.

\bibitem{kwave}
B.~E. Treeby, B.~T. Cox, k-wave: Matlab toolbox for the simulation and reconstruction of photoacoustic wave fields, Journal of biomedical optics 15~(2) (2010) 021314--021314.

\bibitem{speed2001therapeutic}
C.~Speed, Therapeutic ultrasound in soft tissue lesions, Rheumatology 40~(12) (2001) 1331--1336.

\bibitem{softtissue}
M.~O. Culjat, D.~Goldenberg, P.~Tewari, R.~S. Singh, A review of tissue substitutes for ultrasound imaging, Ultrasound in medicine \& biology 36~(6) (2010) 861--873.

\bibitem{thermalprop}
C.~K. McGarry, L.~J. Grattan, A.~M. Ivory, F.~Leek, G.~P. Liney, Y.~Liu, P.~Miloro, R.~Rai, A.~P. Robinson, A.~J. Shih, et~al., Tissue mimicking materials for imaging and therapy phantoms: a review, Physics in Medicine \& Biology 65~(23) (2020) 23TR01.

\bibitem{10.1088/1361-6463/ad5452}
C.~Cengiz, S.~Shahab, \href{http://iopscience.iop.org/article/10.1088/1361-6463/ad5452}{Holographic thermal mapping in volumes using acoustic lenses}, Journal of Physics D: Applied Physics (2024).
\newline\urlprefix\url{http://iopscience.iop.org/article/10.1088/1361-6463/ad5452}

\bibitem{treeby2010k}
B.~E. Treeby, B.~T. Cox, k-wave: Matlab toolbox for the simulation and reconstruction of photoacoustic wave fields, Journal of biomedical optics 15~(2) (2010) 021314--021314.

\bibitem{tsang2013novel}
P.~W.~M. Tsang, T.-C. Poon, Novel method for converting digital fresnel hologram to phase-only hologram based on bidirectional error diffusion, Optics express 21~(20) (2013) 23680--23686.

\bibitem{liu2021pattern}
K.~Liu, Z.~He, L.~Cao, Pattern-adaptive error diffusion algorithm for improved phase-only hologram generation, Chinese Optics Letters 19~(5) (2021) 050501.

\bibitem{ozisik2018inverse}
M.~N. Ozisik, Inverse heat transfer: fundamentals and applications, Routledge, 2018.

\bibitem{taler1999solution}
J.~Taler, W.~Zima, Solution of inverse heat conduction problems using control volume approach, International Journal of Heat and Mass Transfer 42~(6) (1999) 1123--1140.

\bibitem{weber1981analysis}
C.~F. Weber, Analysis and solution of the ill-posed inverse heat conduction problem, International Journal of Heat and Mass Transfer 24~(11) (1981) 1783--1792.

\bibitem{alifanov2012inverse}
O.~M. Alifanov, Inverse heat transfer problems, Springer Science \& Business Media, 2012.

\bibitem{su2004inverse}
J.~Su, G.~F. Hewitt, Inverse heat conduction problem of estimating time-varying heat transfer coefficient, Numerical Heat Transfer, Part A: Applications 45~(8) (2004) 777--789.

\bibitem{huang1999three}
C.-H. Huang, S.-P. Wang, A three-dimensional inverse heat conduction problem in estimating surface heat flux by conjugate gradient method, International Journal of Heat and Mass Transfer 42~(18) (1999) 3387--3403.

\bibitem{perakis2019inverse}
N.~Perakis, O.~J. Haidn, Inverse heat transfer method applied to capacitively cooled rocket thrust chambers, International Journal of Heat and Mass Transfer 131 (2019) 150--166.

\bibitem{haghighi2008two}
M.~G. Haghighi, M.~Eghtesad, P.~Malekzadeh, D.~Necsulescu, Two-dimensional inverse heat transfer analysis of functionally graded materials in estimating time-dependent surface heat flux, Numerical Heat Transfer, Part A: Applications 54~(7) (2008) 744--762.

\bibitem{lee2013inverse}
H.-L. Lee, T.-H. Lai, W.-L. Chen, Y.-C. Yang, An inverse hyperbolic heat conduction problem in estimating surface heat flux of a living skin tissue, Applied Mathematical Modelling 37~(5) (2013) 2630--2643.

\bibitem{mehrabanian2023new}
K.~Mehrabanian, A.~A. Nejad, A new approach for the heat source estimation in cancerous tissue treatment with hyperthermia, International Journal of Thermal Sciences 194 (2023) 108593.

\bibitem{paruch2007identification}
M.~Paruch, E.~Majchrzak, Identification of tumor region parameters using evolutionary algorithm and multiple reciprocity boundary element method, Engineering Applications of Artificial Intelligence 20~(5) (2007) 647--655.

\bibitem{majchrzak2011identification}
E.~Majchrzak, M.~Paruch, Identification of electromagnetic field parameters assuring the cancer destruction during hyperthermia treatment, Inverse Problems in Science and Engineering 19~(1) (2011) 45--58.

\bibitem{hossain2016thermogram}
S.~Hossain, M.~Abdelaal, F.~A. Mohammadi, Thermogram assessment for tumor parameter estimation considering body geometry, Canadian Journal of Electrical and Computer Engineering 39~(3) (2016) 219--234.

\bibitem{scott2018inverse}
E.~P. Scott, Inverse heat transfer for biomedical applications, Theory and Applications of Heat Transfer in Humans 1 (2018) 133--152.

\bibitem{loulou2002thermal}
T.~Loulou, E.~P. Scott, Thermal dose optimization in hyperthermia treatments by using the conjugate gradient method, Numerical Heat Transfer: Part A: Applications 42~(7) (2002) 661--683.

\bibitem{kuznetsov2006optimization}
A.~Kuznetsov, Optimization problems for bioheat equation, International Communications in Heat and Mass Transfer 33~(5) (2006) 537--543.

\bibitem{baghban2015source}
M.~Baghban, M.~B. Ayani, Source term prediction in a multilayer tissue during hyperthermia, Journal of Thermal Biology 52 (2015) 187--191.

\bibitem{liu2020optimized}
X.~Liu, M.~Almekkawy, An optimized control approach for hifu tissue ablation using pde constrained optimization method, IEEE transactions on ultrasonics, ferroelectrics, and frequency control 68~(5) (2020) 1555--1568.

\bibitem{lin2021deep}
Q.~Lin, J.~Wang, F.~Cai, R.~Zhang, D.~Zhao, X.~Xia, J.~Wang, H.~Zheng, A deep learning approach for the fast generation of acoustic holograms, The Journal of the Acoustical Society of America 149~(4) (2021) 2312--2322.

\bibitem{lee2022deep}
M.~H. Lee, H.~M. Lew, S.~Youn, T.~Kim, J.~Y. Hwang, Deep learning-based framework for fast and accurate acoustic hologram generation, IEEE Transactions on Ultrasonics, Ferroelectrics, and Frequency Control 69~(12) (2022) 3353--3366.

\bibitem{li2022acoustic}
B.~Li, M.~Lu, C.~Liu, X.~Liu, D.~Ta, Acoustic hologram reconstruction with unsupervised neural network, Frontiers in Materials 9 (2022) 916527.

\bibitem{goodfellow2014generative}
I.~Goodfellow, J.~Pouget-Abadie, M.~Mirza, B.~Xu, D.~Warde-Farley, S.~Ozair, A.~Courville, Y.~Bengio, Generative adversarial nets, Advances in neural information processing systems 27 (2014).

\bibitem{creswell2018generative}
A.~Creswell, T.~White, V.~Dumoulin, K.~Arulkumaran, B.~Sengupta, A.~A. Bharath, Generative adversarial networks: An overview, IEEE signal processing magazine 35~(1) (2018) 53--65.

\bibitem{durgadevi2021generative}
M.~Durgadevi, et~al., Generative adversarial network (gan): a general review on different variants of gan and applications, in: 2021 6th International Conference on Communication and Electronics Systems (ICCES), IEEE, 2021, pp. 1--8.

\bibitem{mirza2014conditional}
M.~Mirza, S.~Osindero, Conditional generative adversarial nets, arXiv preprint arXiv:1411.1784 (2014).

\bibitem{isola2017image}
P.~Isola, J.-Y. Zhu, T.~Zhou, A.~A. Efros, Image-to-image translation with conditional adversarial networks, in: Proceedings of the IEEE conference on computer vision and pattern recognition, 2017, pp. 1125--1134.

\bibitem{chen2016infogan}
X.~Chen, Y.~Duan, R.~Houthooft, J.~Schulman, I.~Sutskever, P.~Abbeel, Infogan: Interpretable representation learning by information maximizing generative adversarial nets, Advances in neural information processing systems 29 (2016).

\bibitem{karras2019style}
T.~Karras, S.~Laine, T.~Aila, A style-based generator architecture for generative adversarial networks, in: Proceedings of the IEEE/CVF conference on computer vision and pattern recognition, 2019, pp. 4401--4410.

\bibitem{zhu2017toward}
J.-Y. Zhu, R.~Zhang, D.~Pathak, T.~Darrell, A.~A. Efros, O.~Wang, E.~Shechtman, Toward multimodal image-to-image translation, Advances in neural information processing systems 30 (2017).

\bibitem{zhong2024real}
C.~Zhong, Q.~Lu, T.~Li, H.~Su, S.~Liu, Real-time acoustic holography with physics-reinforced contrastive learning for acoustic field reconstruction, Journal of Applied Physics 135~(1) (2024).

\bibitem{xu2023programmable}
M.~Xu, J.~Wang, W.~S. Harley, P.~V. Lee, D.~J. Collins, Programmable acoustic holography using medium-sound-speed modulation, Advanced Science 10~(23) (2023) 2301489.

\bibitem{gu2024holographic}
W.~Gu, J.~Wang, S.~Chai, T.~N. Tran, D.~Ta, X.~Liu, Holographic reconstruction with all-acoustic diffractive network, IEEE Transactions on Computational Imaging (2024).

\bibitem{zhu2009thermal}
L.~Zhu, T.~Schappeler, C.~Cordero-Tumangday, A.~Rosengart, Thermal interactions between blood and tissue: Development of a theoretical approach in predicting body temperature during blood cooling and rewarming, in: Advances in Numerical Heat Transfer, Volume 3, CRC Press, 2009, pp. 209--232.

\bibitem{bredlau2016thermal}
A.-L. Bredlau, M.~McCrackin, A.~Motamarry, K.~Helke, C.~Chen, A.-M. Broome, D.~Haemmerich, Thermal therapy approaches for treatment of brain tumors in animals and humans, Critical Reviews™ in Biomedical Engineering 44~(6) (2016).

\bibitem{coluccia2014first}
D.~Coluccia, J.~Fandino, L.~Schwyzer, R.~O’Gorman, L.~Remonda, J.~Anon, E.~Martin, B.~Werner, First noninvasive thermal ablation of a brain tumor with mr-guided focusedultrasound, Journal of therapeutic ultrasound 2 (2014) 1--7.

\bibitem{bazazian2022perceptually}
D.~Bazazian, B.~Magland, C.~Grimm, E.~Chambers, K.~Leonard, Perceptually grounded quantification of 2d shape complexity, The Visual Computer 38~(9) (2022) 3351--3363.

\bibitem{page2003shape}
D.~L. Page, A.~F. Koschan, S.~R. Sukumar, B.~Roui-Abidi, M.~A. Abidi, Shape analysis algorithm based on information theory, in: Proceedings 2003 international conference on image processing (Cat. No. 03CH37429), Vol.~1, IEEE, 2003, pp. I--229.

\bibitem{chen2005estimating}
Y.~Chen, H.~Sundaram, Estimating complexity of 2d shapes, in: 2005 IEEE 7th Workshop on Multimedia Signal Processing, IEEE, 2005, pp. 1--4.

\bibitem{legon2014transcranial}
W.~Legon, T.~F. Sato, A.~Opitz, J.~Mueller, A.~Barbour, A.~Williams, W.~J. Tyler, Transcranial focused ultrasound modulates the activity of primary somatosensory cortex in humans, Nature neuroscience 17~(2) (2014) 322--329.

\bibitem{legon2022noninvasive}
W.~Legon, A.~Strohman, A.~In, B.~Payne, Noninvasive neuromodulation of subregions of the human insula differentially affect pain processing and heart-rate variability: a within-subjects pseudo-randomized trial, Pain (2022) 10--1097.

\bibitem{strohman2024low}
A.~Strohman, B.~Payne, A.~In, K.~Stebbins, W.~Legon, Low-intensity focused ultrasound to the human dorsal anterior cingulate attenuates acute pain perception and autonomic responses, Journal of Neuroscience 44~(8) (2024).

\bibitem{xu2021histotripsy}
Z.~Xu, T.~L. Hall, E.~Vlaisavljevich, F.~T. Lee~Jr, Histotripsy: the first noninvasive, non-ionizing, non-thermal ablation technique based on ultrasound, International journal of hyperthermia 38~(1) (2021) 561--575.

\bibitem{meng2021applications}
Y.~Meng, K.~Hynynen, N.~Lipsman, Applications of focused ultrasound in the brain: from thermoablation to drug delivery, Nature Reviews Neurology 17~(1) (2021) 7--22.

\bibitem{rabut2020ultrasound}
C.~Rabut, S.~Yoo, R.~C. Hurt, Z.~Jin, H.~Li, H.~Guo, B.~Ling, M.~G. Shapiro, Ultrasound technologies for imaging and modulating neural activity, Neuron 108~(1) (2020) 93--110.

\bibitem{izadifar2020introduction}
Z.~Izadifar, Z.~Izadifar, D.~Chapman, P.~Babyn, An introduction to high intensity focused ultrasound: systematic review on principles, devices, and clinical applications, Journal of clinical medicine 9~(2) (2020) 460.

\bibitem{FUS}
F.~U. Foundation, \href{https://www.fusfoundation.org/the-foundation/foundation-reports/state-of-the-field-report-2023/}{State of the field report}, Tech. rep., Focused ultrasound foundation (October, 2023).
\newline\urlprefix\url{https://www.fusfoundation.org/the-foundation/foundation-reports/state-of-the-field-report-2023/}

\bibitem{ebbini2015ultrasound}
E.~S. Ebbini, G.~Ter~Haar, Ultrasound-guided therapeutic focused ultrasound: Current status and future directions, International journal of hyperthermia 31~(2) (2015) 77--89.

\bibitem{zhou2011high}
Y.-F. Zhou, High intensity focused ultrasound in clinical tumor ablation, World journal of clinical oncology 2~(1) (2011) 8.

\bibitem{partanen2012mild}
A.~Partanen, P.~S. Yarmolenko, A.~Viitala, S.~Appanaboyina, D.~Haemmerich, A.~Ranjan, G.~Jacobs, D.~Woods, J.~Enholm, B.~J. Wood, et~al., Mild hyperthermia with magnetic resonance-guided high-intensity focused ultrasound for applications in drug delivery, International journal of hyperthermia 28~(4) (2012) 320--336.

\bibitem{sammartino2021thermal}
F.~Sammartino, J.~Snell, M.~Eames, V.~Krishna, Thermal neuromodulation with focused ultrasound: implications for the technique of subthreshold testing, Neurosurgery 89~(4) (2021) 610--616.

\end{thebibliography}

\end{document}